\documentclass[12pt]{article}
\usepackage{amsfonts}
\usepackage{amsmath}
\usepackage{amssymb}
\usepackage{amsthm}

\usepackage[utf8]{inputenc}
\usepackage[margin=2cm]{geometry}
\usepackage{geometry}
\usepackage{hyperref}
\usepackage{graphicx}
\usepackage{amsmath}
\usepackage{cite}
\usepackage{hyperref}
\usepackage{tensor}
\usepackage{cases}
\usepackage{appendix}
\usepackage{multirow}
\usepackage{booktabs}
\usepackage{color}
\usepackage{ulem}
\usepackage[font=small]{subfig}
\usepackage[skip=8pt,labelfont={bf,sc},textfont=small]{caption}

\def\[#1\]{\begin{align}#1\end{align}}
\newcommand{\imineq}[2]{\vcenter{\hbox{\includegraphics[height=#2ex]{#1}}}}
\def \nn {\nonumber}
\def \a {\alpha}
\def \b {\beta}

\def \dd{\mathrm{d}}

\def \e {\epsilon}

\def \ve {\varepsilon}

\def \ve{\varepsilon}

\def \s{\sigma}
\def \t{\tau}
\def \ra{\rangle}
\def \la{\langle}
\def \k{\kappa}
\def \mo{\mathcal{O}}

\def \Id{\mathbb{I}}
\def \tC{\tilde{C}}

\def \tr{\text{Tr}}
\def \Tr{\text{Tr}}
\def \T{\mathcal{T}}
\def \N{\mathcal{N}}
\def \Re{\text{Re}}
\begin{document}
\begin{titlepage}
\vspace{0.5cm}
\begin{center}
{\Large \bf {On the real-time evolution of pseudo-entropy in 2d CFTs}}
\lineskip .75em
\vskip 2.5cm
{\large Wu-zhong Guo$^{b,}$\footnote{wuzhong@hust.edu.cn}, Song He$^{a,c,}$\footnote{hesong@jlu.edu.cn; corresponding author},  Yu-Xuan Zhang$^{a,}$\footnote{yuxuanz18@mails.jlu.edu.cn; corresponding author}}
\vskip 2.5em
{\normalsize\it $^{a}$Center for Theoretical Physics and College of Physics, Jilin University,\\ Changchun 130012, People's Republic of China\\
 $^{b}$School of Physics, Huazhong University of Science and Technology\\
Luoyu Road 1037, Wuhan, Hubei 430074, China\\
$^{c}$Max Planck Institute for Gravitational Physics (Albert Einstein Institute),\\
Am M\"uhlenberg 1, 14476 Golm, Germany
}
\vskip 3.0em
\end{center}
\begin{abstract}
In this work, we study the real-time evolution of pseudo-(R\'enyi) entropy, a generalization of entanglement entropy, in two-dimensional conformal field theories (CFTs). We focus on states obtained by acting primary operators located at different space points or their linear combinations on the vacuum. We show the similarities and differences between the pseudo-(R\'enyi) entropy and entanglement entropy. For excitation by a single primary operator, we analyze the behaviors of the 2nd pseudo-R\'enyi entropy in various limits and find some symmetries associated with the subsystem and the positions of the inserted operators. For excitation by linear combinations, the late time limit of the $n$th pseudo-R\'enyi entropy shows a simple form related to the coefficients of the combinations and R\'enyi entropy of the operators, which can be derived by using the Schmidt decomposition. Further, we find two kinds of particular spatial configurations of insertion operators in one of which the pseudo-(R\'enyi) entropy remains real throughout the time evolution.
\end{abstract}
\end{titlepage}

\baselineskip=0.7cm

\tableofcontents
\section{Introduction}
Entanglement entropy, a physical observable stemming from quantum information, has now pervaded many branches of theoretical physics such as quantum many-body physics \cite{Vidal:2002rm,Kitaev:2005dm,PhysRevLett.96.110405}, high energy physics \cite{Bombelli:1986rw,Srednicki:1993im,Calabrese:2004eu,Casini:2009sr,Nishioka:2018khk}, gravitational physics \cite{Ryu:2006bv,Hubeny:2007xt,Swingle:2009bg,Maldacena:2013xja,Almheiri:2020cfm}, and so on. It is worth mentioning that recently, its quantum corrected holographic version \cite{Ryu:2006bv,Hubeny:2007xt}, a.k.a. quantum extremal surface \cite{Engelhardt:2014gca}, in the AdS/CFT correspondence \cite{Maldacena:1997re,Gubser:1998bc,Witten:1998qj} has played a crucial role in solving problems such as black hole information loss \cite{Hawking:1976ra,Mathur:2009hf}, providing a reliable way to further understand quantum gravity.

Also recently, a new quantity associated with a bulk minimal area surface, called pseudo-entropy, has been introduced in \cite{Nakata:2020luh} under the framework of AdS/CFT correspondence. Given a total system $\mathcal{S}$, one can define the pseudo-entropy associated with a subsystem $A$ in terms of the corresponding pseudo-R\'enyi entropy
\[
S_{A}^{(n)}\equiv\frac{1}{1-n}\log\tr\big[(\T_{A}^{1|2})^n\big],\quad(n>1,~n\in\mathbb{Z})\label{pseudo renyi entropy},
\]
where the matrix $\T_{A}^{1|2}$ involved two nonorthogonal quantum states $|\psi_1\ra,|\psi_2\ra\in\mathcal{H}_{\mathcal{S}}$, named reduced transition matrix,  is the partial trace of the transition matrix $\T^{1|2}$,\footnote{$A^c$ refer to the complement of $A$ and we have assumed that the Hilbert space of total system $\mathcal{H}_{\mathcal{S}}$ can be divided into $\mathcal{H}_A\bigotimes\mathcal{H}_{A^c}$.}
\[
\mathcal{T}^{1|2}\equiv\frac{|\psi_1\ra\la\psi_2|}{\la\psi_2|\psi_1\ra},\quad\T_{A}^{1|2}\equiv\tr_{A^c}\big[\mathcal{T}^{1|2}\big]\label{defination of transition matrix}.
\]
The pseudo-entropy of subsystem $A$
\[
S_{A}\equiv-\tr\big[\T_{A}^{1|2}\log\T_A^{1|2}\big]
\]
is obtained by taking the limit of $n\rightarrow1$ for $S_{A}^{(n)}$.  The reduced transition matrix is not Hermitian in general; hence the pseudo-entropy usually takes complex values. However, the results in the qubit system suggest that the real part of pseudo-entropy can be used to characterize the number of distillable Bell pairs averaged over the histories between the initial and final state \cite{Nakata:2020luh}. More intriguingly, it was recently found in \cite{Akal:2021dqt} that the real-time evolution of the real part of pseudo entropy follows the Page curve \cite{Page:1993df} under some field-theoretic settings based on the black hole final state proposal \cite{Horowitz:2003he}. \footnote{More recently, the authors of \cite{Akal:2020twv} reproduce the Page curve in another purely field-theoretic way called moving mirror.}  Hence pseudo-entropy, like entanglement entropy, can reflect certain underlying correlation structures. Refer to \cite{Mollabashi:2020yie,Camilo:2021dtt,Mollabashi:2021xsd,Nishioka:2021cxe,Goto:2021kln,Miyaji:2021lcq,Mukherjee:2022jac,Ishiyama:2022odv} for other related developments of pseudo-entropy.

The main purpose of this paper is to study the real-time evolution of the real part of pseudo-entropy for locally excited state generated by a single primary operator or linear combination of a bunch of them in various 2d CFTs. Unlike the case in \cite{Akal:2021dqt}, our investigation can be regarded  as a pseudo-entropy extension of the real-time evolution of the entanglement entropy after such local operator excitations \cite{Nozaki:2014hna}. In recent years, the time evolution of entanglement entropy for locally  excited states has been widely studied, including rational CFTs \cite{He:2014mwa,Caputa:2015tua,Chen:2015usa}, irrational CFTs \cite{He:2017lrg}, large-$c$ CFTs \cite{Caputa:2014vaa,Caputa:2016tgt}, boundary CFTs \cite{Guo:2015uwa}, warped CFTs \cite{Apolo:2018oqv}, CFTs at finite temperature \cite{Caputa:2014eta}, multiple local excitations \cite{Guo:2018lqq}, and holographic duals of the local excitations \cite{Nozaki:2013wia,Ageev:2021qsr,Zenoni:2022eka,Kawamoto:2022etl}. In 2d rational CFTs (RCFTs), it was found that the variation of $n$th R\'enyi entropy for locally primary excited states saturates to a constant equal to the logarithm of the quantum dimension of the local operator’s conformal family \cite{He:2014mwa,Caputa:2015tua,Chen:2015usa}. Such saturation get well interpreted in the picture of propagation of quasiparticles pairs \cite{Calabrese:2005in,Nozaki:2014hna}. On the other hand, it was found in large-$c$ CFTs\cite{Caputa:2014vaa,Caputa:2016tgt} that a characteristic feature called scrambling of entanglement would scramble the information of non-perturbative constants like quantum dimensions and lead to a logarithmically diverged  R\'enyi entropy \cite{Caputa:2014vaa,Asplund:2015eha}.

Since pseudo-entropy is a straightforward generalized concept of entanglement entropy, we shall study the time evolution behavior of  pseudo-entropy for locally excited states in various 2d CFTs. We set up various situations to calculate the pseudo-entropy of locally excited states in 2d rational CFTs and large-$c$ CFTs and to explore universal properties of pseudo entropy of locally excited states.

This paper is organized as follows. Section \ref{sec2} outlines the standard replica method to compute the $n$th pseudo-R\'enyi entropy for locally excited states. In Section \ref{sec3} we mainly focus on the case of $n = 2$. We first study the limiting behaviors of real-time evolution of $2$nd pseudo-R\'enyi entropy in rational CFTs and large-$c$  CFTs. We then numerically analyze the full-time evolutions of $2$nd pseudo-R\'enyi entropy in some specific interacting theories.   In Section \ref{sec4}, we extend the above analysis to the $n$th pseudo-R\'enyi entropy. We end in Section \ref{sec5} with conclusions and prospects. Some useful formulae and calculation details are presented in the appendices.
\section{General calculations of pseudo-R\'enyi entropy}\label{sec2}
\subsection{Setup for local excitations and $S_{A}^{(n)}$  from replica method }
In this section, we review the replica calculation for the pseudo-R\'enyi entropy \cite{Nakata:2020luh}, which is almost same with that for the ordinary R\'enyi entropy \cite{Calabrese:2004eu}. Consider a 2d CFT dwells on a plane $\Sigma_1$ with coordinates $\{\t,~x\}$ ($\dd s^2=\dd \t^2+\dd x^2$). We are primarily interested in the cases that $|\psi_{1}\ra$, $|\psi_{2}\ra$ are states defined by acting various operators on the ground state $|\Omega\ra$,
\[
|\psi_{j}\ra=\frac{1}{\sqrt{\mathcal{N}_{j}}}\mo_{j,1}(-\t_{j,1},x_{j,1})\mo_{j,2}(-\t_{j,2},x_{j,2})...\mo_{j,n_{j}}(-\t_{j,n_{j}},x_{j,n_{j}})|\Omega\ra&,\label{chap2:psi12}\\
(\t_{j,i+1}\geq\t_{j,i}>0,~i=1,2,...,n_{j}-1;~j=1,2),&\nn
\]
where $\N_{j}$ is normalization factor and $\mo_{j,i}(-\t_{j,i},x_{j,i})\equiv e^{-H\t_{j,i}}\mo_{j,i}(x_{j,i})e^{H\t_{j,i}}$ is operator located at $(\t=-\t_{j,i}, x=x_{j,i})$. We can write down the  corresponding transition matrix \eqref{defination of transition matrix} in terms of the path integral language as follows
\[
\mathcal{T}^{1|2}=&\frac{\mo_{1,1}(-\t_{1,1},x_{1,1})...\mo_{1,n_{1}}(-\t_{1,n_1},x_{1,n_1})|\Omega\ra\la\Omega|\mo^\dagger_{2,n_{2}}(\t_{2,n_2},x_{2,n_2})...\mo^{\dagger}_{2,1}(\t_{2,1},x_{2,1})}{\la\Omega|\mo^\dagger_{2,n_{2}}(\t_{2,n_2},x_{2,n_2})...\mo^{\dagger}_{2,1}(\t_{2,1},x_{2,1})\mo_{1,1}(-\t_{1,1},x_{1,1})...\mo_{1,n_{1}}(-\t_{1,n_1},x_{1,n_1})|\Omega\ra}\nn\\[.08cm]
=&\left(\imineq{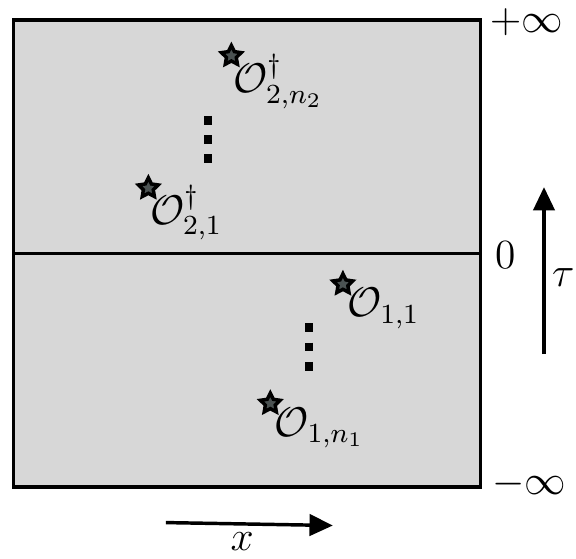}{17}\right)^{-1}\times~~~\imineq{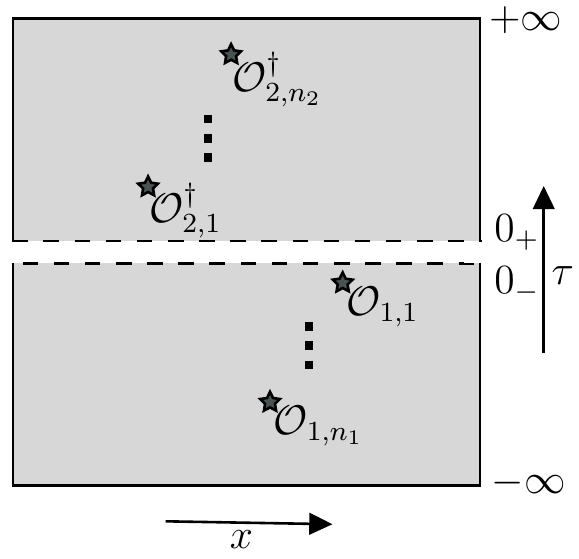}{17}.
\]
Here the dashed lines represent the free boundaries, and the stars denote the insertion points of the operators. The reduced transition matrix of the subsystem $A$ is obtained by "stitching up" the upper and lower edges of $A^c$
\[
\mathcal{T}_A^{1|2}=\tr_{A^c}\big[\mathcal{T}^{1|2}\big]=&\left(\imineq{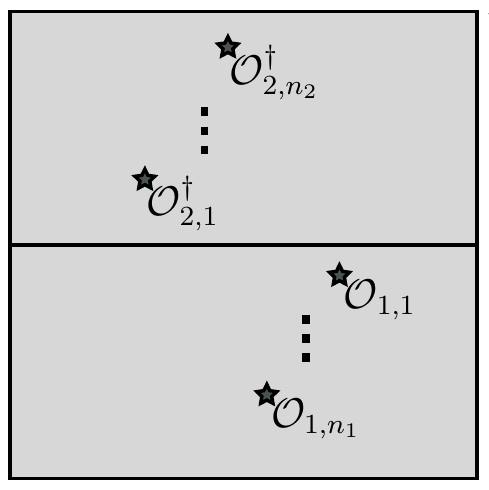}{17}\right)^{-1}\times~~~\imineq{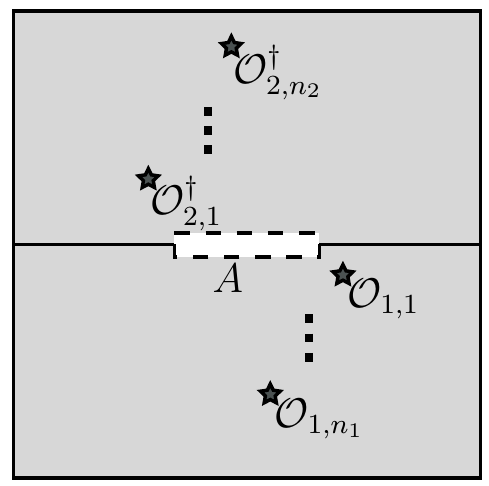}{17}.\label{reduced transition matrix A: graph}
\]
Substituting \eqref{reduced transition matrix A: graph} into \eqref{pseudo renyi entropy}, the path integral representation of $S_{A}^{(n)}$ is given by
\[
S_{A}^{(n)}=&\frac{1}{1-n}\log\left[\left(\imineq{reducedtransitionpart1.pdf}{17}\right)^{-n}\times~~~\imineq{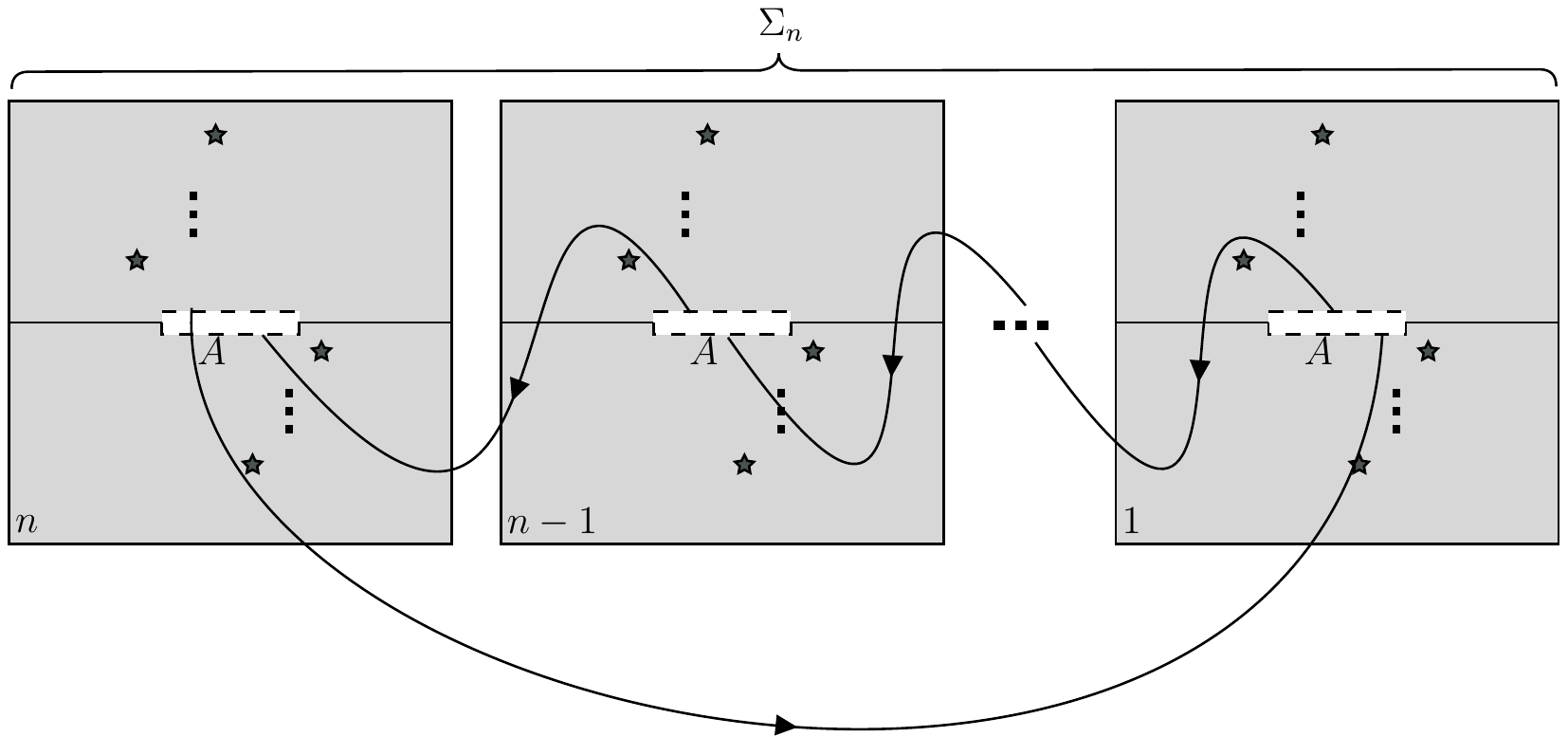}{25}\right]\nn\\
=&S_{A;vac}^{(n)}+\frac{1}{1-n}\left(\log\big\la(\mo_{2,n_2}^{\dagger(n)}...\mo^{(n)}_{1,n_1})...(\mo_{2,n_2}^{\dagger(1)}...\mo^{(1)}_{1,n_1})\big\ra_{\Sigma_n}-n\log\big\la\mo_{2,n_2}^{\dagger}...\mo_{1,n_1}\big\ra_{\Sigma_1}\right).\label{mthPEEinpathintegtrallanguage}
\]
In the above, $S^{(n)}_{A;vac}$ stands for the $n$th R\'enyi entropy of $A$ when the total system is in vacuum state, and $\Sigma_n$ is a $n$-sheeted Riemann surface constructed by gluing $n$ sheets $\Sigma_1$ together at subsystem $A$. The subscript of $\mo^{(k)}$ denotes that the operator $\mo$ is living on the $k$th sheet of $\Sigma_n$. Since  $S^{(n)}_{A;vac}$ does not carry any information about excitations, we shall focus on the excess  $\Delta S_{A}^{(n)}\equiv S_{A}^{(n)}-S_{A;vac}^{(n)}$ hereafter.
\subsection{The excess of second pseudo-R\'enyi entropy $\Delta S_{A}^{(2)}$}
\begin{figure}[htbp]
  \centering
  \includegraphics[width=0.4\linewidth]{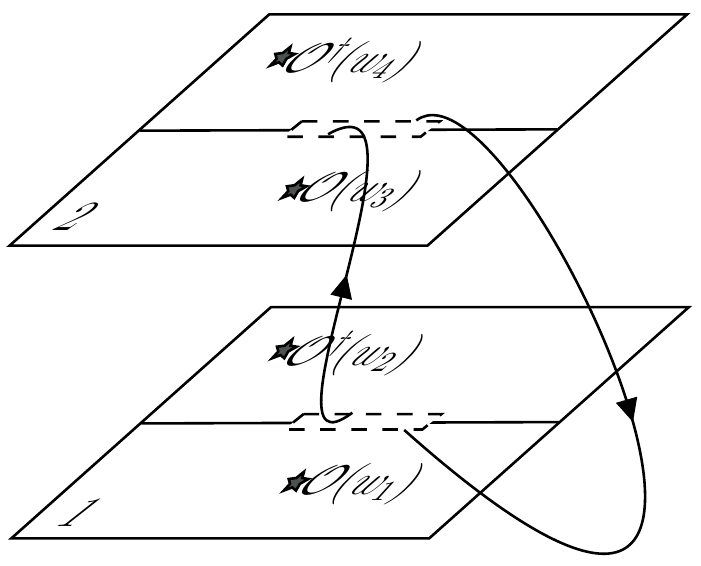}
  \caption{The 2-sheeted space $\Sigma_2$. The dashed box represents the subsystem $A$. }\label{sigma2}
\end{figure}

Let us first concentrate on the simplest case that $n_{1}=n_2=1,~n=2$. In the meantime, we are mainly interested in the case where two inserted operators are the same.
Now \eqref{chap2:psi12} can be reduced to
\[
|\psi_j\ra=\frac{1}{\sqrt{\mathcal{N}_j}}\mo(-\t_j,x_j)|\Omega\ra,\quad(j=1,2),\label{chap2:realpsi12}
\]
and the corresponding excess of pseudo-R\'enyi entropy is given by
\[
\Delta S_A^{(2)}=-\log\frac{\big\la\mo^{\dagger(2)}(\t_{2},x_{2})\mo^{(2)}(-\t_{1},x_{1})\mo^{\dagger(1)}(\t_{2},x_{2})\mo^{(1)}(-\t_{1},x_{1})\big\ra_{\Sigma_2}}{\big\la\mo^{\dagger}(\t_{2},x_{2})\mo(-\t_{1},x_{1})\big\ra_{\Sigma_1}^2}.\label{2thPEE}
\]
The above expression is reduced to two- and four-point functions that we know for precisely solvable CFTs. It coincides with the excess of the second R\'enyi entropy when the insertion points are symmetric about the $x$-axis, i.e., $\t_{1}=\t_{2}$ and $x_{1}=x_{2}$.
Owing to the conformal symmetry, when $\mo$ is a primary operator with chiral and antichiral conformal dimension $\Delta_\mo$, the two-point and four-point function of $\mo$ on $\Sigma_1$ are given by\footnote{$\{z,\bar{z}\}:=\{x+i\t,x-i\t\}$ are complex coordinates on $\Sigma_1$, likewise for complex coordinates $\{w,\bar{w}\}$  on $\Sigma_n$. $z_{ij}\equiv z_i-z_j$.}
\[
\la\mo^{\dagger}(z_2,\bar{z}_2)\mo(z_1,\bar{z}_1)\ra_{\Sigma_1}&=\frac{c_{12}}{|z_{12}|^{4\Delta_\mo}},\label{2pt function on sigma1}\\
\la\mo^{\dagger}(z_4,\bar{z}_4)\mo(z_3,\bar{z}_3)\mo^{\dagger}(z_2,\bar{z}_2)\mo(z_1,\bar{z}_1)\ra_{\Sigma_1}&=|z_{13}z_{24}|^{-4\Delta_\mo}G(\eta,\bar{\eta}),\label{4ptandG}
\]
respectively, where  $(\eta,\bar{\eta})\equiv\big(z_{12}z_{34}/(z_{13}z_{24}),~\bar{z}_{12}\bar{z}_{34}/(\bar{z}_{13}\bar{z}_{24})\big)$ are cross ratios and $c_{12}$ is normalization factor. Since we can apply the conformal transformation
\[
z=&\left(\frac{w}{w-L}\right)^{1/n},\quad(A=[0,L]),\label{map1}\\
z=&w^{1/n},~~~~~~~~~~~~\quad(A=[0,+\infty)),\label{map2}
\]
to map $\Sigma_n$ to $\Sigma_1$, we obtain  the four-point function on $\Sigma_2$ by applying the above conformal maps with $n=2$
\begin{equation}
\la\mo^{\dagger}(w_4,\bar{w}_4)\mo(w_3,\bar{w}_3)\mo^{\dagger}(w_2,\bar{w}_2)\mo(w_1,\bar{w}_1)\ra_{\Sigma_2}=
\begin{cases}
\big|\frac{16L^2z_1^2z_2^2}{(z_1^2-1)^2(z_2^2-1)^2}\big|^{-4\Delta_\mo}G(\eta,\bar{\eta})~&(A=[0,L]),\\
\\
\big|16z_1^2z_2^2\big|^{-4\Delta_\mo}G(\eta,\bar{\eta})~&(A=[0,+\infty))\label{4pt on sigma2},
\end{cases}
\end{equation}
where we have set
\[
(w_3,\bar{w}_3)_{\text{sheet~2}}&=(w_1,\bar{w}_1)_{\text{sheet~1}}=(x_1-i\t_1,x_1+i\t_1),\nn\\
(w_4,\bar{w}_4)_{\text{sheet~2}}&=(w_2,\bar{w}_2)_{\text{sheet~1}}=(x_2+i\t_2,x_2-i\t_2),
\]
as shown in figure \ref{sigma2}. 
Combining \eqref{4pt on sigma2} with \eqref{2pt function on sigma1}, the excess of the second pseudo-R\'enyi entropy \eqref{2thPEE} is expressed as a function which depends only on $\eta$ and $\bar{\eta}$
\[
\Delta S^{(2)}_{A}=\log\frac{c_{12}^2}{\big|\eta(1-\eta)\big|^{4\Delta_\mo}\cdot G(\eta,\bar{\eta})}.\label{2thpee}
\]

\section{Real-time evolution of $\Re[\Delta S_A^{(2)}]$}\label{sec3}
The real-time evolution of the pseudo-R\'enyi entropy for locally excited states can be regarded as a generalization of that of the R\'enyi entropy for locally excited states \cite{Nozaki:2014hna, He:2014mwa, Nozaki:2014uaa,Caputa:2014vaa, Caputa:2014eta,Guo:2015uwa, Caputa:2015tua,Chen:2015usa, He:2017lrg}. In rational CFTs, it is known that the excess of the R\'enyi entropy saturates to a constant equal to the logarithm of the quantum dimension of the inserted primary operator\cite{Nozaki:2014hna, He:2014mwa}. A similar result for pseudo-R\'enyi entropy is found in free CFT in \cite{Nakata:2020luh}. However, \cite{Nakata:2020luh} only consider the real-time evolution of pseudo-R\'enyi entropy with different insertion time and the same insertion spatial coordinates. Richer evolutionary structures seem to lie in another insertion configuration of operators---two operators with different spatial coordinates and the same insertion time. In this section, we mainly explore this insertion configuration and give a general argument in the light of \cite{He:2014mwa}.
\subsection{$\Delta S_A^{(2)}$ for two primary operators with different spatial coordinates}
In the following, we explore the case where the time coordinates of the two inserted operators are the same, but the spatial coordinates are different. That is, we are considering the following real-time dependent transition matrix
\[
\mathcal{T}^{1|2}(t)\equiv\frac{e^{-iHt}e^{-\e H}\mo(x_1)|\Omega\ra\la\Omega|\mo^{\dagger}(x_2)e^{-\epsilon H}e^{iHt}}{\la\Omega|\mo^{\dagger}(x_2)e^{-2\e H}\mo(x_1)|\Omega\ra},\quad \T_A^{1|2}(t)\equiv\Tr_{A^c}[\T^{1|2}(t)].\label{TforsingleO}
\]
It amounts to perform the following analytic continuation for $\t_{1}$, $\t_2$ in \eqref{chap2:realpsi12}:
\[
\t_1=\e+it,\quad\t_2=\e-it,
\]
wherein $\e>0$ is an infinitesimally small regularization parameter to suppress the high energy modes \cite{Calabrese:2005in}.
\subsubsection{Early time, middle time, and late time behaviors}
Let us first study some limiting behaviors of the pseudo-R\'enyi entropy to obtain some generic conclusions. The procedure is similar to that of entanglement entropy\cite{He:2014mwa,Caputa:2014vaa}. For the subsystem $A$ of infinite length, we are mainly interested in the early time and late time limits of pseudo-R\'enyi entropy, while for the subsystem $A$ of finite length, we are also interested in the middle time limit.\footnote{Based on the result of entanglement entropy in finite scales \cite{He:2014mwa}, the middle time may be defined as the interval $[u,v]$, where $u=\min\big[|x_1|,|x_2|,|L-x_1|,|L-x_2|\big]$, $v=\max\big[|x_1|,|x_2|,|L-x_1|,|L-x_2|\big]$.}
\paragraph{\textbf{The subsystem} $A=[0,\infty)$:} Consider the case of $A=[0,\infty)$, in which we are mainly interested in the early ($t\rightarrow0$) and late ($t\rightarrow+\infty$) time limits. According to the expression of the $2$nd pseudo-R\'enyi entropy \eqref{2thpee}, it's helpful to study the early and late time behaviors of $\eta$ and $\bar{\eta}$ firstly, which we summarize in the table \ref{tab:eta}.\footnote{The details of derivation can be found in appendix \ref{appen-a1}.}
\begin{table}[hbtp]
\centering
\caption{Early time and late time behaviors of $(\eta,\bar{\eta})$ for the subsystem $A=[0,\infty)$.}
\label{tab:eta}
\begin{tabular}{|cl|cl|cc|}
\hline
\hline
\multicolumn{2}{|c|}{\multirow{2}{*}{$(\eta,\bar{\eta})$}} & \multicolumn{2}{c|}{\multirow{2}{*}{$x_1x_2>0$}} & \multicolumn{2}{c|}{\multirow{2}{*}{$x_1x_2<0$}}                        \\
\multicolumn{2}{|c|}{}                     & \multicolumn{2}{c|}{}                     & \multicolumn{2}{c|}{}                                            \\ \hline
\multicolumn{2}{|l|}{\multirow{2}{*}{Late time $(t\to\infty)$}} & \multicolumn{2}{c|}{\multirow{2}{*}{$(1,0)$}} & \multicolumn{2}{c|}{\multirow{2}{*}{$(1,0)$}}                        \\
\multicolumn{2}{|c|}{}                     & \multicolumn{2}{c|}{}                     & \multicolumn{2}{c|}{}                                            \\ \hline
\multicolumn{2}{|l|}{\multirow{4}{*}{Early time $(t\to0)$}} & \multicolumn{2}{c|}{} & \multicolumn{1}{c|}{\multirow{2}{*}{$x_1>0>x_2$}} & \multirow{2}{*}{$x_2>0>x_1$} \\
\multicolumn{2}{|c|}{}                     & \multicolumn{2}{c|}{\multirow{1}{*}{$(\frac{1}{2}+a,\frac{1}{2}+a)$}}                     & \multicolumn{1}{c|}{}                     &                      \\ \cline{5-6}
\multicolumn{2}{|c|}{}                     & \multicolumn{2}{c|}{\multirow{2}{*}{$a=\frac{x_1+x_2}{4\sqrt{x_1x_2}}$}}                     & \multicolumn{1}{c|}{\multirow{2}{*}{$\big(\frac{1}{2}+a,\frac{1}{2}-a\big)$}} & \multirow{2}{*}{$\big(\frac{1}{2}-a,\frac{1}{2}+a\big)$} \\
\multicolumn{2}{|c|}{}                     & \multicolumn{2}{c|}{}                     & \multicolumn{1}{c|}{}                     &                      \\ \hline
\end{tabular}
\end{table}
We can see from the table that for general space configurations of the two inserted points, the late time behaviors of cross ratios are uniform. Still, the early time behavior is somewhat intricate. One can, however, obtain concise results by thinking about the situations where two operators are very close together. Intuitively, we may expect the results to degenerate into the case of R\'enyi entropy. The quadratic limit of cross ratios is given by
\begin{align}
&\lim\limits_{x_1\rightarrow x_2}\lim\limits_{t\rightarrow0}(\eta,\bar{\eta})\simeq
\begin{cases}
(0,0),&x_2<0\\
(1,1),&x_2>0\label{infA+early:x1<x2}
\end{cases},
\end{align}
These coincide with the second R\'enyi entropy results in \cite{Chen:2015usa}.
Another intriguing case is to set $x_2=-x_1\neq0$, where  the early time limit of cross ratios is reduced to
\[
\lim\limits_{t\rightarrow0}(\eta,\bar{\eta})\simeq\big(\frac{1}{2},\frac{1}{2}\big).\label{limitanalysis:x1+x2=0}
\]
We next follow the arguments in \cite{He:2014mwa,Caputa:2014vaa} to cope with the function $G_(\eta,\bar{\eta})$ in \eqref{2thpee}. In general CFTs,  $G(\eta,\bar{\eta})$ can be expressed as follows using the conformal blocks \cite{Belavin:1984vu}
\[
G(\eta,\bar{\eta})=\sum_{p}(C^p_{\mo^{\dagger}\mo})^2F_\mo(p|\eta)\bar{F}_\mo(p|\bar{\eta}),\label{conformalblockexpand}
\]
where $C_{\mo^{\dagger}\mo}^p$ is the coefficient of the three-point function $\la\mo^{\dagger}\mo\phi_p\ra$ and the index $p$ corresponds to each $\phi_p$ of all Virasoro primary fields. We can normalize $G(\eta,\bar{\eta})$ such that the two-point function \eqref{2pt function on sigma1} has a unit normalization $c_{12}= 1$, and it leads to the following behavior of $F_\mo(p|\eta)$ in $\eta\rightarrow0$
limit\footnote{Note we set $\phi_0$ to be equal to the identity operator, which means $\Delta_0=0$. }
\[
\lim\limits_{\eta\rightarrow0}F_\mo(p|\eta)=\eta^{\Delta_p-2\Delta_\mo}(1+O(\eta)).\label{f0}
\]
The above behavior indicates that as $\eta$ goes to zero, the identity operator dominates the contribution in the summation of \eqref{conformalblockexpand}. Moreover, with the bootstrap relation
\[
G(\eta,\bar{\eta})=G(1-\eta,1-\bar{\eta}),\label{bootstrap}
\]
we obtain  behaviors of $G(\eta,\bar{\eta})$ in two limits for general 2d CFTs
\[
\lim\limits_{\eta\to0\atop \bar{\eta}\to 0} G(\eta,\bar{\eta})\simeq|\eta|^{-4\Delta_\mo},\quad \lim\limits_{\eta\to1\atop \bar{\eta}\to 1}G(\eta,\bar{\eta})\simeq|1-\eta|^{-4\Delta_\mo}.\label{early for infA}
\]
The above results correspond precisely to the early time behavior of the $G(\eta,\bar\eta)$ function when the spatial positions of the two inserted operators tend to coincide.

The late time behavior of $G(\eta,\bar{\eta})$, according to the table \ref{tab:eta},  requires some knowledge of the behavior of conformal block in the limit $\eta\to1$. For rational CFTs,  the fusion transformation \cite{Moore:1988uz,Moore:1988ss}
\[
F_\mo(p|1-\eta)=\sum_q F_{pq}[\mo]\cdot F_\mo(q|\eta)\label{fusionransformation}
\]
can be exploited to  give the expression of $F_\mo(p|\eta)$ in $\eta\to 1$ limit and thus fix the leading-order of the late time behavior of $G$
\[
\lim_{\eta\to1\atop\bar{\eta}\to0}G(\eta,\bar{\eta})\simeq d_\mo^{-1}(1-\eta)^{-2\Delta_\mo}\bar{\eta}^{2\Delta_\mo},\label{latercftinfA}
\]
where $d_\mo=1/F_{00}[\mo]$ is the quantum dimension \cite{Moore:1988ss} of $\mo$. Combine \eqref{2thPEE} with \eqref{early for infA} and \eqref{latercftinfA}, we find the following expression for the second pseudo-R\'enyi entropy
\[
\Delta S_{A}^{(2)}\simeq
\begin{cases}
0,&t\to0~\&\&~x_1\sim x_2,\\
\log d_{\mo},&t\to\infty.\label{peefprrcft}
\end{cases}
\]
It is noted that the late limit of both $2$nd pseudo-R\'enyi entropy and R\'enyi entropy saturates to $\log d_{\mo}$, which indicates that the quasiparticle pair picture seems to be preserved in the pseudo-entropy.

We next move to analyse large-$c$ CFTs. For large-$c$ CFTs, in the limit $\Delta_p/c\ll1$, the conformal block has the following universal form \cite{Fateev:2011qa,Zamolodchikov:1984eqp}
\[
F_{\mo}(p,\eta)\simeq \eta^{\Delta_p-2\Delta_\mo} \cdot\mathbin{_2F_1}(\Delta_p,\Delta_p,2\Delta_p;\eta),\label{f1}
\]
where $\mathbin{_2F_1}(a,b,c;z)$ is the   hypergeometric function. Whereas, the authors in \cite{Caputa:2014vaa} argued that the above approximation fails when $\eta$ is very close to $1$ such as   $|1-\eta|\sim (D^p_\mo)^{-\frac{1}{2\Delta_\mo}}$, where $D^p_\mo\sim \exp(c^\alpha)$ ($\alpha$ is a certain positive constant) is exponentially large and in RCFTs $D_\mo^0$ coincides with the quantum dimension $d_\mo$. When $\eta$ is very close to $1$, the leading order of $F_{\mo}(p,\eta)$ is given by \cite{Caputa:2014vaa}
\[
F_{\mo}(p,\eta)\simeq\frac{1}{D^p_\mo}\cdot(1-\eta)^{-2\Delta_\mo},\quad\big(|1-\eta|\lesssim(D^p_\mo)^{-\frac{1}{2\Delta_\mo}}\big).
\]
Furthermore, following the arguments in \cite{Caputa:2014vaa}, the summation in Eq.\eqref{conformalblockexpand} can be approximated by counting the contribution of the conformal vacuum block
\[
G(\eta,\bar{\eta})\simeq F_\mo(0,\eta)\bar{F}_\mo(0,\bar{\eta})\label{gff}
\]
when we are considering large-$c$ theroies with gravity duals.
Eq.(\ref{f0}) with Eq.(\ref{f1}-\ref{gff}) together lead to another type of the late time limit of $G(\eta,\bar{\eta})$
\[G(\eta,\bar{\eta})\simeq
\begin{cases}
\bar{\eta}^{-2\Delta_\mo},& |1-\eta|\gtrsim(D^p_\mo)^{-\frac{1}{2\Delta_\mo}},\\
\frac{1}{D^0_\mo}\cdot(1-\eta)^{-2\Delta_\mo}\bar{\eta}^{-2\Delta_\mo},&|1-\eta|\lesssim(D^p_\mo)^{-\frac{1}{2\Delta_\mo}}.\label{latelargecinfA}
\end{cases}
\]
Substituting \eqref{early for infA} and \eqref{latelargecinfA} into \eqref{2thpee} respectively, We obtain three distinct stages of evolution of the second pseudo-R\'enyi entropy.
\[
\Delta S_{A}^{(2)}\simeq
\begin{cases}
0,& t\to0~\text{and}~x_1\sim x_2,\\
\log\big[\big(-(\frac{x_1-x_2+2i\e}{4t})^2\big)^{-2\Delta_\mo}\big],&\max\big[|x_1|,|x_2|,\e\big]\ll t\lesssim\frac{1}{4}(D^0_\mo)^\frac{1}{4\Delta_\mo}\sqrt{(x_1-x_2)^2+4\e^2},\\
\log D^0_\mo,&t\gtrsim\frac{1}{4}(D^0_\mo)^\frac{1}{4\Delta_\mo}\sqrt{(x_1-x_2)^2+4\e^2}.
\end{cases}
\]
Like R\'enyi entropy \cite{Caputa:2014vaa}, the second pseudo-R\'enyi entropy has an intermediate process of logarithmic time evolution.
If we take the real part of $\Delta S_A^{(2)}$, as we are interested in, the corresponding logarithm evolves as follows
\[
\text{Re}\big[\Delta S_A^{(2)}\big]&=4\Delta_{\mo}\log\frac{4t}{\sqrt{(x_1-x_2)^2+4\e^2}},
\]
which matches the result in \cite{Caputa:2014vaa} when two space points coincide. It's rewarding to mention that when we take the large-$c$ limit first, like in holography, $D_\mo^{0}$ will go to infinity, and $\Delta S_{A}^{(2)}$ is left with logarithmical growth.

\paragraph{The subsystem $A=[0,L]$:} In the case of finite scale, the early time and late time behaviors of the cross ratios  can be readily obtained from the expressions of cross ratios \eqref{appA2,etabafterac} after the analytic continuation,
\begin{align}
\lim_{t\to0}(\eta,\bar{\eta})\simeq&
\begin{cases}
(\frac{1}{2}+b,\frac{1}{2}+b),&x_1,x_2<0||0<x_1,x_2<L||L<x_1,x_2,\\
(\frac{1}{2}+b,\frac{1}{2}-b),&x_1>L>x_2>0||L>x_1>0>x_2,\\
(\frac{1}{2}-b,\frac{1}{2}+b),&x_2>L>x_1>0||L>x_2>0>x_1,\\
(\frac{1}{2}-b,\frac{1}{2}-b),&x_1>L>0>x_2||x_2>L>0>x_1,\quad\Big(b=\frac{L(x_1+x_2)-2x_1x_2}{4\sqrt{x_1x_2(L-x_1)(L-x_2)}}\Big),\label{a=l:early11}
\end{cases}\\
\lim_{t\to\infty}(\eta,\bar{\eta})\simeq&\Big(-\frac{L^2(x_2-x_1+2i\epsilon)^2}{16t^4},-\frac{L^2(x_2-x_1-2i\epsilon)^2}{16t^4}\Big)\simeq(0,0).\label{a=l:late}
\end{align}
Once again, we encounter a complicated early time behavior, and it can be simplified by taking quadratic limit $(x_1\sim x_2~\text{and}~t\sim0)$
\[
\lim_{x_1\to x_2}\lim_{t\to0}(\eta,\bar{\eta})\simeq
\begin{cases}
(0,0),&x_2<0||L<x_2,\\
(1,1),&0<x_2<L.\label{a=l:early}
\end{cases}
\]
Furthermore, we find that another interesting class of space configurations, $|x_1|,|x_2|\gg L$, can also reduce the early time results \eqref{a=l:early11},
\[\lim_{t\to0}(\eta,\bar{\eta})\simeq
(0,0),\quad(|x_1|,|x_2|\gg L).
\]

To analytically extract the middle time $\big(t\in[u,v]$, $u=\min\big[|x_1|,|x_2|,|L-x_1|,|L-x_2|\big]$, $v=\max\big[|x_1|,|x_2|,|L-x_1|,|L-x_2|\big]\big)$ behavior of cross ratios, let us consider the large $L$ limit, that is,  $L\gg|x_1-x_2|$.  This is because one can expect that when $L\gg|x_1-x_2|$, the middle time behavior of $\Delta S_A^{(2)}$ will tend to the late time behavior of $\Delta S_A^{(2)}$ of the infinite subsystem. Consider two special spatial configurations that satisfy the constraint: i. $L\gg\max\big[|x_1|,|x_2|\big]$; ii. $L\gg\max\big[|L-x_1|,|L-x_2|\big]$. The above two configurations correspond to the situation
where
operators live concentratedly near the left and right boundaries of $A$, respectively.\footnote{One may also interested in the opposing situation that the operators are scattered at both ends of $A$. The middle time behavior of cross ratios in this case is found to be
\[\lim_{t\to\frac{\sqrt{2}L}{2}}(\eta,\bar{\eta})\simeq
\Big(\frac{1}{2}+O(1/L),\frac{1}{2}+O(1/L)\Big),\quad L\gg\max\big[|x_1|,|L-x_2|\big].\label{bothendlimit}
\]
} In these cases, the value of the cross ratios at a typical middle time, $t=L/2$, can be calculated analytically
\[
\lim_{t\to\frac{L}{2}}(\eta,\bar{\eta})\simeq
\begin{cases}
\left(1+\frac{(x_2-x_1+2i\epsilon)^2}{L^2},-\frac{(x_2-x_1-2i\epsilon)^2}{9L^2}\right)\simeq(1,0),&{L\gg \max\big[|x_1|,|x_2|\big]},\\
\\
\left(-\frac{(x_2-x_1+2i\epsilon)^2}{9L^2},1+\frac{(x_2-x_1-2i\epsilon)^2}{L^2}\right)\simeq(0,1),&{L\gg \max\big[|L-x_1|,|L-x_2|\big]}.
\end{cases}
\]
We obtain a middle-time behavior similar to the late-time behavior \eqref{app1:latetimeeta} for the infinite subsystem. For more general cases, the numerical calculation shows that
\[
(\eta,\bar{\eta})\Big\lvert_{t\in[u',v']}\simeq
\begin{cases}
(1,0),&L\gg|x_1-x_2|~\&\&~x_{\text{min}}<\frac{L-|x_1-x_2|}{2},\\
(0,1),&L\gg|x_1-x_2|~\&\&~x_{\text{min}}>\frac{L-|x_1-x_2|}{2},\quad(x_{\text{min}}\equiv\min\big[x_1,x_2\big]),\label{a=l:middle}
\end{cases}
\]
where $u'=\min\big[\big\{|x_1|,|x_2|,|L-x_1|,|L-x_2|\big\}\setminus\{u\}\big]$, $v'=\max\big[\big\{|x_1|,|x_2|,|L-x_1|,|L-x_2|\big\}\setminus\{v\}\big]$ for $x_1\neq x_2$ and $[u',v']=[u,v]$ for $x_1=x_2$.

Combining with the previous discussion of $G(\eta,\bar{\eta})$, according to Eqs.\eqref{a=l:early}, \eqref{a=l:middle}, and \eqref{a=l:late}, we get the picture of the evolution of $\Delta S_{A}^{(2)}$ under some constraints in RCFTs\footnote{Note that normally the quantum dimension $d_\mo$ is real, and this is true in all the models considered in this paper (i.e. we have $\bar{d}_\mo=d_\mo$ in this paper).}
\[
\Delta S_{A}^{(2)}\simeq
\begin{cases}
0,&t\sim0~\&\&~\big(|x_1-x_2|\sim 0~||~L\ll|x_1|,|x_2|\big),\\
\log d_\mo,&t\in[u',v']~\&\&~L\gg|x_1-x_2|~\&\&~x_{\min}<\frac{L-|x_1-x_2|}{2},\\
\log \bar{d}_\mo,&t\in[u',v']~\&\&~L\gg|x_1-x_2|~\&\&~x_{\min}>\frac{L-|x_1-x_2|}{2},\\
0,&t\to\infty.\label{SA=Llimit}
\end{cases}
\]
\subsubsection{Examples in 2d CFTs}
In the previous subsection, we have studied several limiting behaviors of the second pseudo-R\'enyi entropy. However, there are still some mysteries about the evolution of $\Delta S_A^{(2)}$ that limit analysis is infeasible to solve:

\begin{enumerate}
  \item The intermediate process of the evolution of $\Re[\Delta S_A^{(2)}]$ from an initial value to $\log d_\mo$ in RCFTs.
  \item Is there anything special about $\text{Re}[\Delta S_A^{(2)}]$ evolution in certain symmetric spatial configurations (such as $x_1=-x_2$ for $A=[0,\infty)$)?
\end{enumerate}

In this subsection, we will resort to numerical analysis to uncover the whole time evolution picture of $\Delta S_A^{(2)}$ under several specific 2d CFT models. We expect that the above problems will be answered to some extent in these concrete models. Before entering into the numerical study, we first point out some model-independent symmetries of the second pseudo-R\'enyi entropy, which are reflected in the following examples.
\paragraph{Symmetries for $\Delta S_A^{(2)}$:}
Re-examining  the cross ratios of finite and infinite subsystem (\eqref{appA2,etabafterac} and \eqref{appA1,eta},respectively), one can find some hidden symmetries of them,
\begin{align}
&\eta(x_2,x_1,t)=\big[\eta(x_1,x_2,t)\big]^*,\quad \bar{\eta}(x_2,x_1,t)=\big[\bar{\eta}(x_1,x_2,t)\big]^*,\label{jiaohuan12}\\
&\eta(-x_1,-x_2,t)=1-\bar{\eta}(x_1,x_2,t),\quad\quad\quad \big(A=[0,\infty)\big),\label{jiafuhao}\\
&\eta(L-x_1,L-x_2,t)=\bar{\eta}(x_1,x_2,t),\quad \quad\quad\big(A=[0,L]\big),\label{Lduicheng}
\end{align}
where  $"*"$ denotes complex conjugate. Further, it's easy to show that the above symmetries may be extended  to $\Delta S_A^{(2)}$ when the $G(\eta,\bar{\eta})$ function has the following properties
\[
G(\eta,\bar{\eta})=&G(\bar{\eta},\eta),\label{Gxingzhi1}\\
G(\eta^*,\bar{\eta}^*)=&[G(\eta,\bar{\eta})]^*\label{Gxingzhi2}.
\]
Combining \eqref{2thpee}, \eqref{jiaohuan12}, and \eqref{Gxingzhi2}, one  obtains the first symmetry  of $\Delta S_A^{(2)}$, 
\[
\Delta S_A^{(2)}(x_2,x_1,t)=
\begin{cases}
\Delta S_A^{(2)}(x_1,x_2,t),&\tr\big[\big(\T_A^{1|2}(t)\big)^2\big]\in\mathbb{R}_-,\\
\big[\Delta S_A^{(2)}(x_1,x_2,t)\big]^*,&\tr\big[\big(\T_A^{1|2}(t)\big)^2\big]\in\mathbb{C}\setminus\mathbb{R}_-.\label{sym1forSA}
\end{cases}
\]
Combining \eqref{2thpee}, \eqref{bootstrap}, and (\ref{jiafuhao}-\ref{Gxingzhi1}), the second symmetry of   $\Delta S_A^{(2)}$ reads
\[
\Delta S_{[0,L]}^{(2)}(x_1,x_2,t)=&\Delta S_{[0,L]}^{(2)}(L-x_1,L-x_2,t),\label{SA=fini}\\
\Delta S_{[0,\infty)}^{(2)}(x_1,x_2,t)=&\Delta S_{[0,\infty)}^{(2)}(-x_1,-x_2,t).\label{SA=INF}
\]
There are some physical or mathematical understandings that may explain the appearance  of the above symmetries. For $A=[0,L]$,  both $A$ and $A^c$ are invariant under reflection with respect to $x=L/2$, which implies  two sets of space configurations are symmetric. Thus  pseudo-R\'enyi entropy should be equal with the inserted points $(x_1,x_2)$ and $(L-x_1,L-x_2)$, hence we obtain \eqref{SA=fini}; For $A=[0,\infty)$, we may  have the equality $S^{(n)}_A(x_1,x_2,t)=S^{(n)}_{A^c}(-x_1,-x_2,t)$ in terms of the symmetry of the system. In addition to the basic property $S^{(n)}(\T^{1|2}_{A^c})=S^{(n)}(\T^{1|2}_{A})$ of the $n$th pseudo-R\'enyi entropy \cite{Nakata:2020luh}, we obtain \eqref{SA=INF}; Eq.\eqref{sym1forSA} can be interpreted by the fact that exchanging $x_1$ with $x_2$ is equivalent to let $\T^{1|2}(t)\to\big(\T^{1|2}(t)\big)^\dagger$. The above argument suggests that these symmetries hold not only for the $2$nd pseudo-R\'enyi entropy, but also for any order. We shall make a numerical examination on them in section \ref{sec4}.

On the other hand, one can see that there are two special space configurations --- $x_1=L-x_2$ for $A=[0,L]$ and $x_1=-x_2$ for $A=[0,\infty)$,  that are screened out by these  symmetries. Taking $x_1=-x_2,~A=[0,\infty)$ as an example, the operation of swapping $x_1$ and $x_2$ is equivalent to the spatial reflection operation, which means that $\Delta S_{[0,\infty)}^{(2)}(x_1,-x_1,t)$ is real when $\tr\big[\big(\T_{[0,\infty)}^{1|2}(t)\big)^2\big]\in\mathbb{C}\setminus\mathbb{R}_-$. For $A=[0,L]$, simple algebra shows that $\Delta S_{[0,L]}^{(2)}(x_1,L-x_1,t)=\log\frac{c_{12}^2}{|\eta(1-\eta)|^{4\Delta_\mo}G(\eta,\eta^*)}$.\footnote{The absolute value here is $|\eta|\equiv\sqrt{\eta\cdot\eta^*}$.} Since one can expect $G(\eta,\eta^*)$ to be greater than 0, we obtain a real second pseudo-R\'enyi entropy evolution in this insertion configuration, whose correctness is verified in subsequent examples.

Finally, when only paying attention to the real part of $\Delta S^{(2)}_A$, the above results show that the evolution of $\Re[\Delta S_A^{(2)}]$ may be "4-fold degenerate",
\[
&\Re[\Delta S_{[0,L]}^{(2)}(x_1,x_2,t)]=\Re[\Delta S_{[0,L]}^{(2)}(L-x_1,L-x_2,t)]\nn\\
=&\Re[\Delta S_{[0,L]}^{(2)}(x_2,x_1,t)]=\Re[\Delta S_{[0,L]}^{(2)}(L-x_2,L-x_1,t)],\label{degener1}\\
\nn\\
&\Re[\Delta S_{[0,\infty)}^{(2)}(x_1,x_2,t)]=\Re[\Delta S_{[0,\infty)}^{(2)}(-x_1,-x_2,t)]\nn\\
=&\Re[\Delta S_{[0,\infty)}^{(2)}(x_2,x_1,t)]=\Re[\Delta S_{[0,\infty)}^{(2)}(-x_2,-x_1,t)].\label{degener2}
\]
Hence we may choose to label each space configuration with the following parameters
\[
x_m\equiv\frac{x_1+x_2}{2},\quad l\equiv|x_1-x_2|.
\]
\paragraph{Example I--- Free scalar:} Let us warm up with a simple example --- the $c=1$ free scalar, and choose the operator $\mo=\frac{1}{\sqrt{2}}\big(e^{\frac{i}{2}\phi}+e^{-\frac{i}{2}\phi}\big)$ which has (chiral and antichiral) conformal dimension $\Delta_\mo=\frac{1}{8}$ and  quantum dimension $d_{\mo}=2$. The corresponding function $G(\eta,\bar{\eta})$ is found to be $ G(\eta,\bar{\eta})=\frac{1+|\eta|+|1-\eta|}{2\sqrt{|\eta||1-\eta|}}$, which
apparently satisfies Eqs.(\ref{Gxingzhi1}), (\ref{Gxingzhi2}) and gives the following concise expression of $\Delta S_A^{(2)}$,
\[
\Delta S_A^{(2)}(\eta,\bar{\eta})=\log\frac{2}{1+|\eta|+|1-\eta|}.\label{SAinFB}
\]
On the other hand, utilizing the identity $
\big\la\sigma(z_1,\bar{z}_1)...\sigma(z_{2n},\bar{z}_{2n})\big\ra^2_{\Sigma_1}=\big\la \mo(z_1,\bar{z}_1)...\mo(z_{2n},\bar{z}_{2n})\big\ra_{\Sigma_1}$  and $\Delta_\mo=2\Delta_\sigma$ \cite{DiFrancesco:1987ez},
where $\sigma$ is the spin operator in Ising model, it can be found that $\Delta S_{A}^{(n)}[\mo\text{-excitation}]=2\Delta S^{(n)}_{A}[\sigma\text{-excitation}].$ Thus our calculations in this part are also applicable to the
case of $\sigma$-excitation in Ising model.
\begin{figure}[htpb]
	\centering
\captionsetup[subfloat]{farskip=10pt,captionskip=1pt}
\subfloat{
			\includegraphics[width =0.45\linewidth]{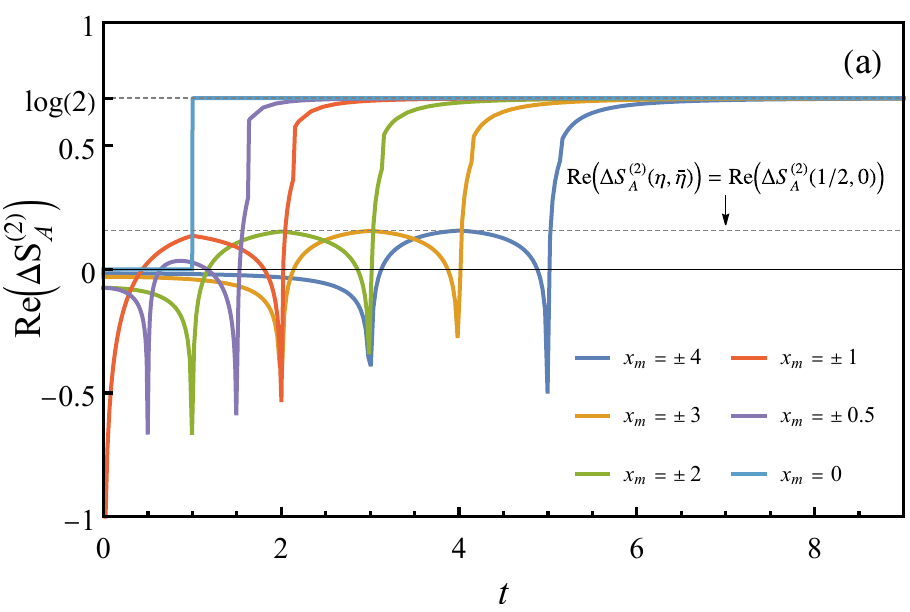}}
\hfill
\subfloat{
			\includegraphics[width =0.45\linewidth]{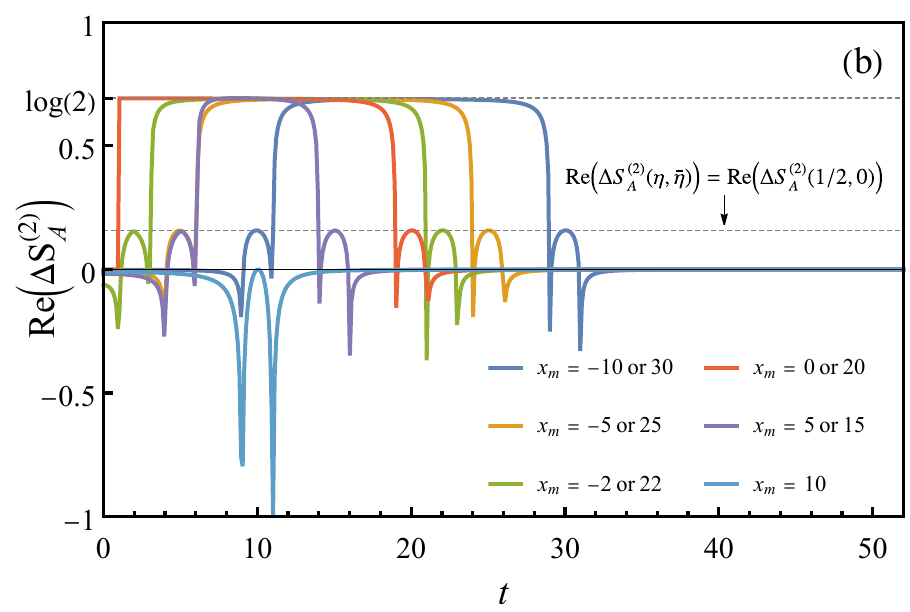}}\\

\subfloat{
			\includegraphics[width =0.45\linewidth]{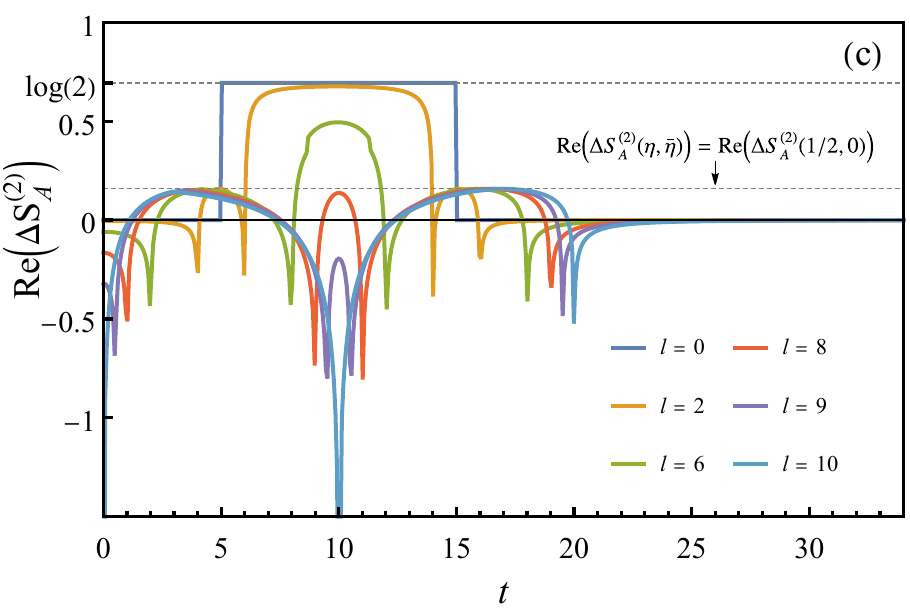}}
\hfill
\subfloat{
			\includegraphics[width =0.45\linewidth]{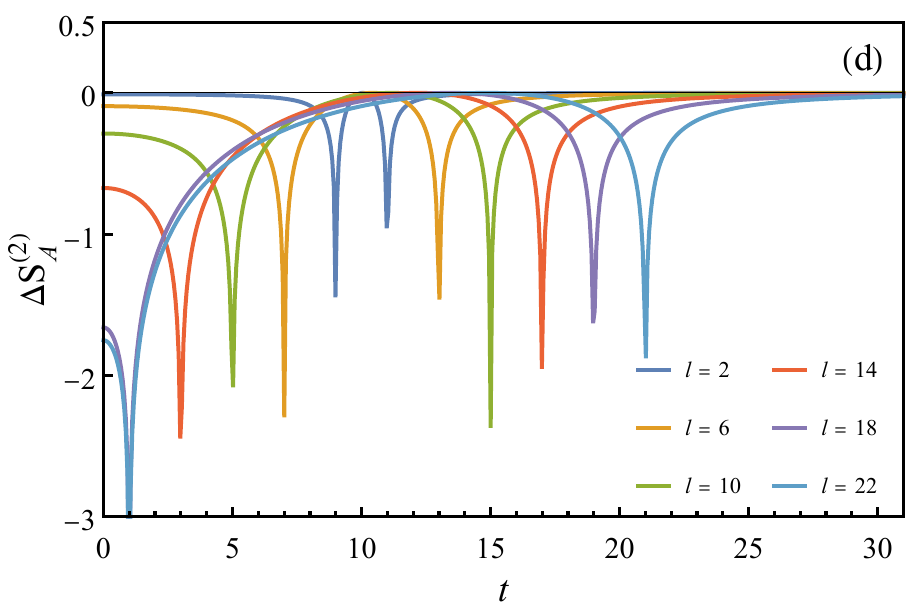}}
	\caption{The real-time evolution of $\Re[\Delta S_A^{(2)}]$ in $c=1$ free scalar, where the insertion operator is chosen to be $\frac{1}{\sqrt{2}}(e^{i\phi}+e^{-i\phi})$ and the regulator $\e=10^{-5}$. (a): $A=[0,\infty)$, $l=2$; (b): $A=[0,20],~l=2$;~(c): $A=[0,10]$, $x_m=-5$;~(d): $A=[0,20]$, $x_m=10$.}
	\label{Fig:FB}
	\vspace{-0.5em}
\end{figure}

Now we turn to the numerical component. The first is the  simpler case that $A=[0,\infty)$. As shown in figure \ref{Fig:FB}(a), since the relative size between the spacing of two operators and the scale of the subsystem is not involved in this case, we fix $l$ ($l=2$) and adjust $x_m$ to observe the evolution of $\Re[S_A^{(2)}]$. One can read a common feature from these evolving curves (except for the case of $x_m=0$): There is a small hump  between the early time evolution determined by \eqref{early determin} and the late time evolution determined by Eq.\eqref{app1:latetimeeta}.  We can explain the appearance of the hump to some extent in terms of the picture of quasiparticle propagation. Under the quasiparticle propagation picture, the time nodes at which the hump shape evolution begins and ends correspond exactly to the time nodes at which two quasiparticle pairs moving at the speed of light from the insertion points enter or leave the subsystem $A$. It can be found that the peak of the hump shape evolution is reached at $t\simeq|x_m|$, and the value of the cross ratios at this time is given by
\begin{align}
\lim_{t\to|x_m|}(\eta,\bar{\eta})=
\begin{cases}
\Big(\frac{1}{2}+O(t-|x_m|),\frac{1}{2}+\frac{x_m}{\sqrt{4x_m^2-\frac{1}{4}l^2+\epsilon^2-i\epsilon l}}+O(t-|x_m|)\Big),&x_m<0,\\
\Big(\frac{1}{2}+\frac{x_m}{\sqrt{4x_m^2-\frac{1}{4}l^2+\epsilon^2-i\epsilon l}}+O(t-|x_m|),\frac{1}{2}+O(t-|x_m|)\Big),&x_m>0.
\end{cases}
\end{align}
Obviously $\frac{x_m}{\sqrt{4x_m^2-\frac{1}{4}l^2+\e^2-i\e l}}\to\pm\frac{1}{2}$ when $|x_m|\gg l$, but the approximation is accurate enough when $|x_m|\geq l$. Thus we have
\[
\lim_{t\to|x_m|}(\eta,\bar{\eta})\simeq
\begin{cases}
\big(\frac{1}{2},0\big),&x_m\leq-l,\\
\big(1,\frac{1}{2}\big),&x_m\geq l.
\end{cases}
\]
However, due to the symmetries of $G(\eta,\bar{\eta})$, the above two cases give the same value of $\Re[S_A^{(2)}]$. For the most special case of $x_m=0$,
which we have already encountered in the limit analysis \eqref{limitanalysis:x1+x2=0}, $\Re[S_A^{(2)}]$ behaves exactly like the second R\'enyi entropy \cite{Nozaki:2014uaa}. In fact, we find $\Delta S_A^{(2)}$ is real in this case,
\[
\Delta S_{[0,\infty)}^{(2)}(x,-x,t)=
\begin{cases}
0,&0\leq t<|x|,\\
\log 2,&t>|x|,
\end{cases}
\]
which is consistent with our previous symmetry analysis.
Next, we categorize the case of $A=[0,L]$ in terms of the relative size between the spacing of two operators and the scale of the subsystem. When the distance between two operators is much smaller than the length of the subsystem (i.e. $l\ll L$), as depicted in figure \ref{Fig:FB}(b), we reproduce the evolution pattern described by Eq.\eqref{SA=Llimit}. Notice that we lost the middle time behavior of $\log d_\mo$ in the case of $x_m=10$, since in this case the related time interval (interval in Eq.\eqref{a=l:middle}) is a null set.  Unlike the case of the infinite subsystem, We can see that the small hump virtually appears twice in the evolving curves in figure \ref{Fig:FB}(b). Again, this coincides with the picture of quasiparticle pairs propagation in a finite  subsystem. The cross ratios for the peaks of the humps are found to be
\begin{align}\lim_{t\to|x_m|}(\eta,\bar{\eta})\simeq&
\begin{cases}
\big(\frac{1}{2},0\big),&L\gg l~\&\&~x_m\leq -l,\\
\big(1,\frac{1}{2}\big),&L\gg l~\&\&~l\leq x_m\leq (L-l)/2,\\
\big(0,\frac{1}{2}\big),&L\gg l~\&\&~(L+l)/2\leq x_m\leq L-l,\\
\big(0,\frac{1}{2}\big),&L\gg l~\&\&~x_m\geq L+l.\\
\end{cases}\\
\lim_{t\to|L-x_m|}(\eta,\bar{\eta})\simeq&
\begin{cases}
\big(\frac{1}{2},0\big),&L\gg l~\&\&~x_m\leq -l,\\
\big(\frac{1}{2},0\big),&L\gg l~\&\&~l\leq x_m\leq (L-l)/2,\\
\big(\frac{1}{2},1\big),&L\gg l~\&\&~(L+l)/2\leq x_m\leq L-l,\\
\big(0,\frac{1}{2}\big),&L\gg l~\&\&~x_m\geq L+l.\\
\end{cases}
\end{align}
$\Re[S_A^{(2)}(\eta,\bar{\eta})]$
is equal in the above cases, on account of the symmetries of $G(\eta,\bar{\eta})$. We then gradually increase the spacing between operators such that the constraint $l\ll L$ no longer holds (see figure \ref{Fig:FB}(c)). We can see that the intermediate behavior of $\log d_\mo$ gradually vanishes as $l$ increases, but the peaks of the humps seem to remain the same. Another interesting case, as we have discussed in symmetry analysis, is to fix $x_m=L/2$ and then gradually increase $l$. As shown in figure \ref{Fig:FB}(d), we do obtain a real pseudo-R\'enyi entropy. Meanwhile, the result shows that  the middle time behavior of $S_A^{(2)}$ tends to zero instead of $\log d_\mo$, and the time to saturation of middle time behavior also shifts from $\frac{1}{2}L$ to $\frac{\sqrt{2}}{2}L$ \eqref{bothendlimit} with the increase of $l$.

\paragraph{Example II---Minimal model:}
Another simple example is the excitation of $(2,1)$ operator $\phi_{(2,1)}$ in the minimal models $\mathcal{M}(p,p')$ with $p>p'$.
The conformal dimension and quantum dimension of the $\phi_{(2,1)}$ are well-known to be $\Delta_{(2,1)}=\frac{3p}{4p'}-\frac{1}{2}$ and $d_{(2,1)}=-2\cos\frac{\pi p}{p'}$, respectively. In addition, it has a relatively simple
four-point function \cite{Dotsenko:1984nm,Dotsenko:1984ad} that satisfies Eqs.(\ref{Gxingzhi1}) and (\ref{Gxingzhi2}),
\[
G(\eta,\bar{\eta})=|\eta|^{\frac{p}{p'}}|1-\eta|^{\frac{p}{p'}}\cdot\left[\frac{\sin\left(\frac{\pi p}{p'}\right)\sin\left(\frac{3\pi p}{p'}\right)}{\sin\left(\frac{2\pi p}{p'}\right)}|I_1(\eta)|^2+\frac{\sin\left(\frac{\pi p}{p'}\right)\sin\left(\frac{\pi p}{p'}\right)}{\sin\left(\frac{2\pi p}{p'}\right)}|I_2(\eta)|^2\right],
\]
where the functions $I_{1,2}$ are defined as follows
\[
I_1(\eta)=&\frac{\Gamma\left(\frac{3p}{p'}-1\right)\Gamma\left(1-\frac{p}{p'}\right)}{\Gamma\left(\frac{2p}{p'}\right)}\cdot \mathbin{_2F_1}\left(\frac{p}{p'},-1+\frac{3p}{p'},\frac{2p}{p'},\eta\right),\nn\\
I_2(\eta)=&\eta^{1-\frac{2p}{p'}}\frac{\Gamma\left(1-\frac{p}{p'}\right)\Gamma\left(1-\frac{p}{p'}\right)}{\Gamma\left(2-\frac{2p}{p'}\right)}\cdot \mathbin{_2F_1}\left(\frac{p}{p'},1-\frac{p}{p'},2-\frac{2p}{p'},\eta\right).
\]
The normalization factor $c_{12}$ in \eqref{2thpee} can be read off by taking the limit $z_{12}=z_{34}\to0$ of the four-point function\footnote{We have $\la\phi_{(2,1)}(z_1,\bar{z}_1)\phi_{(2,1)}(z_2,\bar{z}_2)\phi_{(2,1)}(z_3,\bar{z}_3)\phi_{(2,1)}(z_4,\bar{z}_4)\ra\to c_{12}^2|z_{12}|^{-8\Delta_{(2,1)}}$ in the limit of $z_{12}=z_{34}\to0$.}, and the result turns out to be $
c_{12}^2=\frac{\sin\big(\frac{\pi p}{p'}\big)^2}{\sin\big(\frac{2\pi p}{p'}\big)}\cdot\frac{\Gamma\big(1-\frac{p}{p'}\big)^4}{\Gamma\big(2-\frac{2p}{p'}\big)^2}$. Since the minimal models are unitary iff $p-p'=\pm1$, below we  consider only these unitary cases  and use critical Ising $\mathcal{M}(4,3)$, tricritical Ising $\mathcal{M}(5,4)$, three-state Potts at criticality $\mathcal{M}(6,5)$ and so on as prototypical examples.
\begin{figure}[t]
	\centering
\captionsetup[subfloat]{farskip=5pt,captionskip=1pt}
\subfloat{
			\includegraphics[width =0.45\linewidth]{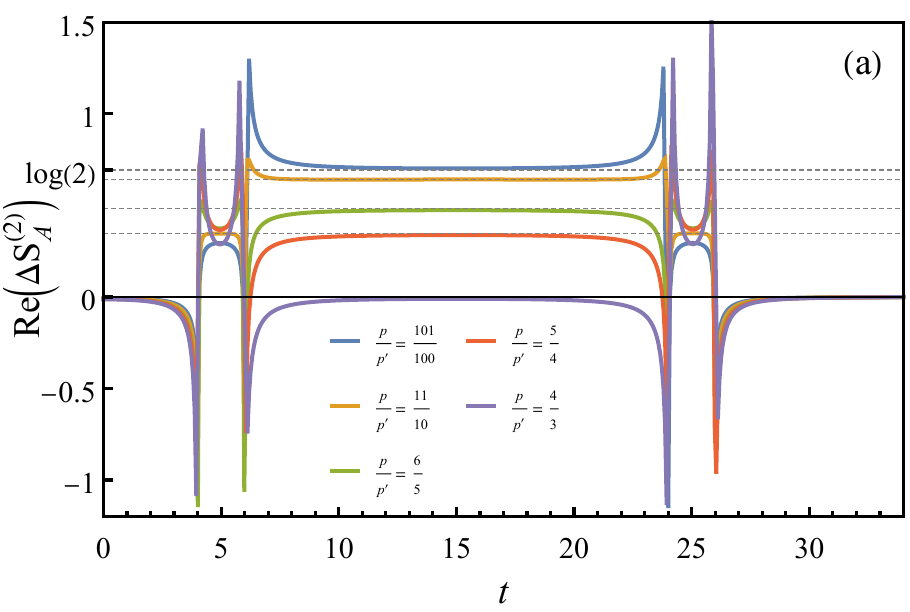}}
\hfill
\subfloat{
			\includegraphics[width =0.45\linewidth]{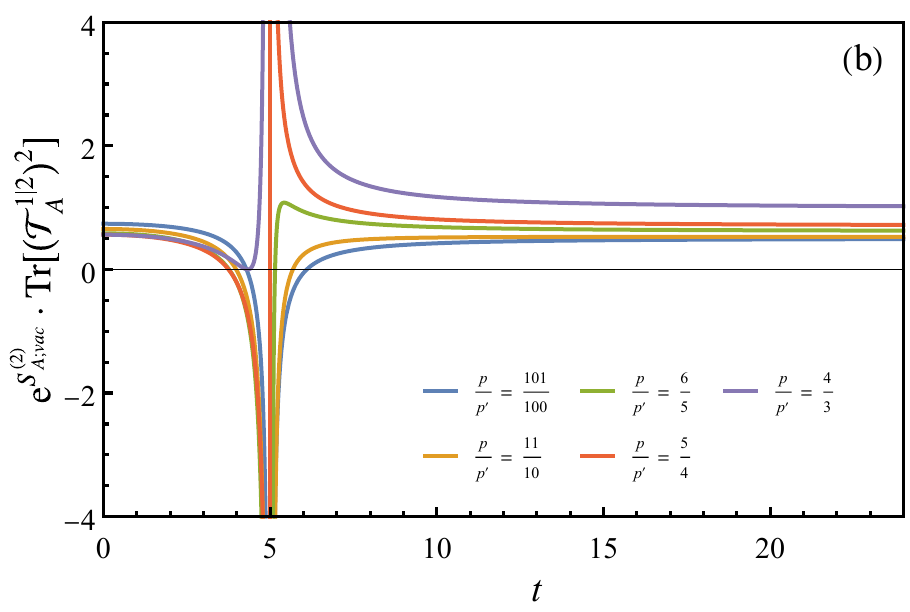}}\\

\subfloat{
			\includegraphics[width =0.45\linewidth]{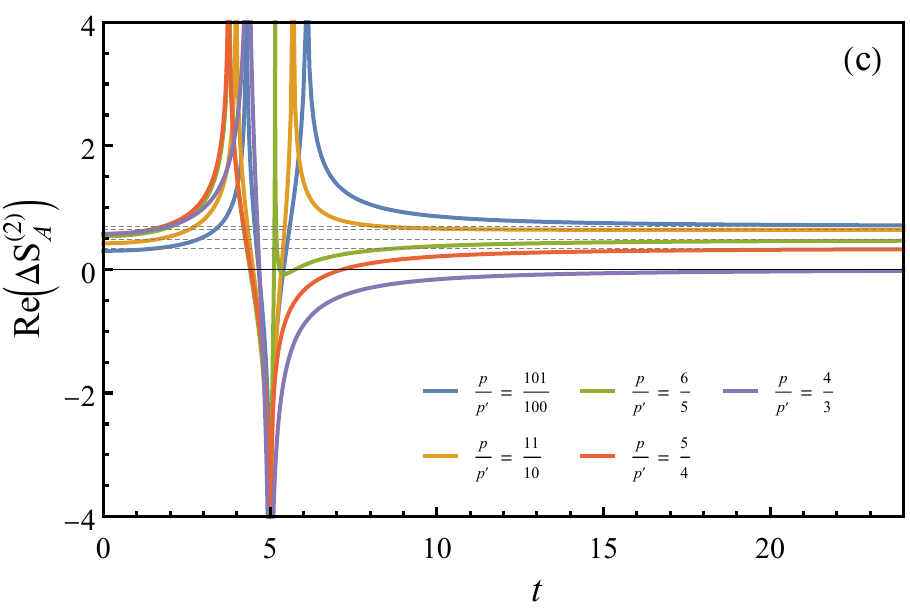}}
\hfill
\subfloat{
			\includegraphics[width =0.45\linewidth]{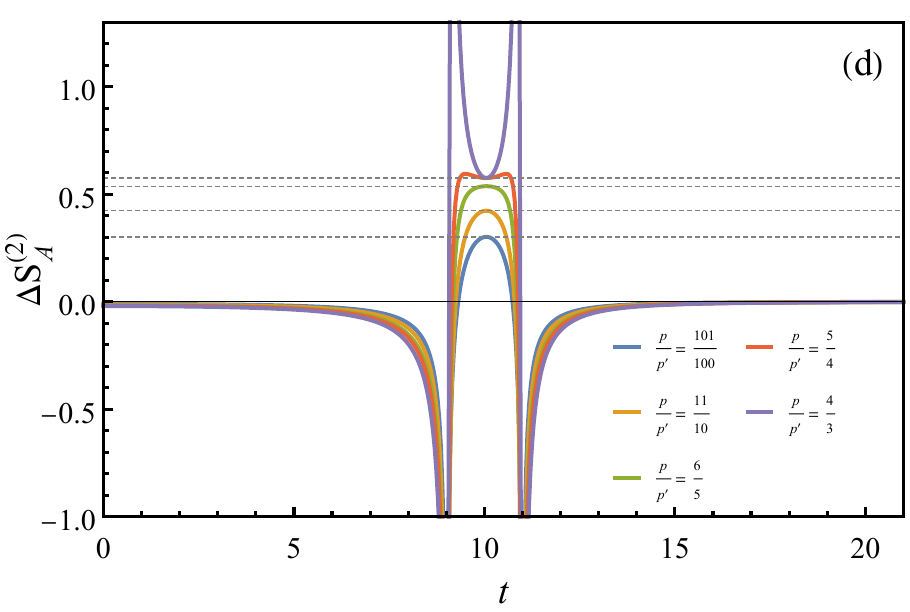}}
	\caption{(a) and (c): The time evolution of $\Re[\Delta S_A^{(2)}]$  under the $\phi_{(2,1)}$-excitation in minimal models. The regulator is chosen to be $\e=10^{-5}$. We have $A=[0,20]$, $l=2$, $x_m=-5$ for (a), and $A=[0,\infty)$, $x_m=0$, $l=10$ for (c). The dashed lines correspond to $\log d_{(2,1)}$ for different $p/p'$; ~(b): The time evolution of $\text{e}^{S^{(2)}_{A;vac}}\cdot\Tr[(\T_A^{1|2})^2]$ in the case of $A=[0,\infty)$, where $S^{(2)}_{A;vac}$ is the 2nd R\'enyi entropy of $A$ when the total system is in the vacuum state. The parameters are selected as $\e=10^{-5}$, $x_m=0$, $l=10$; ~(d): The time evolution of $\Delta S_A^{(2)}$ in the case of $A=[0,L]$. The parameters are selected as $L=20$, $\e=10^{-5}$, $x_m=10$, $l=2$. $\Delta S^{(2)}_A(\eta,\bar\eta)=\Delta S^{(2)}_A(1/2,1/2)$ at the dashed lines.}
	\label{Fig:Mini}
	\vspace{-0.5em}
\end{figure}

Now we turn to the numerical part to observe which properties of the evolution of $\Delta S_A^{(2)}$ in the free scalar are retained in the minimal models. Figure \ref{Fig:Mini}(a) and \ref{Fig:Mini}(c) demonstrate the full-time evolution of $\Re[\Delta S^{(2)}_A]$ in the cases of $A=[0,L]~(L\gg l)$ and $A=[0,\infty)$, respectively. In these two cases, it can be found that $\Re[\Delta S^{(2)}_A]$ saturates to the theoretical value $\log d_{(2,1)}$ in the middle and late time, respectively, which coincides with the case of free scalar. Whereas, since figure $\ref{Fig:Mini}$(c) is drawn under the first symmetric space configuration ( $x_1=-x_2=\pm5$), a comparison
between the corresponding curve ($x_m=0$) in figure \ref{Fig:FB}(a) shows that there are significant differences in the behavior of $\Delta S_A^{(2)}$ in two theories: i). Except for Ising model $\mathcal{M}(4,3)$, $\Delta S_A^{(2)}$ no longer remains real in the full-time evolution, which is manifest to seen in figure $\ref{Fig:Mini}$(b).\footnote{Note that the complex $\Delta S_A^{(2)}$ is not contradictory with the previous symmetry analysis, because by symmetry analysis we can only prove that $\Tr[(\mathcal{T}_A^{1|2})^2]$ is real in the case of $A=[0,\infty]$.} The trace of $(\T^{1|2}_A)^2$ is negative over an interval except for the Ising model, which results in a complex $\Delta S_A^{(2)}$; ii). The evolution of pseudo-R\'enyi entropy in the case of $x_m=0$  no longer behaves like that of R\'enyi entropy; Figure \ref{Fig:Mini}(d) exhibits the evolution of $\Delta S_A^{(2)}$ under the second symmetric space configuration $x_m=L/2$ in the case of $A=[0,L]$. As predicted by the symmetry analysis, we can see that  $\Delta S^{(2)}_A$ is real throughout the time evolution. Notice that $(\eta,\bar\eta)$ takes the value of $(1/2,1/2)$ at the peak or valley of the middle time evolution.

\paragraph{Example III---Wess–Zumino–Witten model:} The last example we would like to explore is the excitation of $g^\a_\b(z,\bar z)$ operator in a Wess–Zumino–Witten (WZW) model with affine Lie algebra $\text{SU}(N)_k$. The operator $g^\a_\b(z,\bar z)$ in the fundamental representation $\a=\{1,0,...,0\}$ has the (chiral and antichiral) conformal dimension $\Delta_g=\frac{N^2-1}{2N\k}$ and quantum dimension $d_g=N^{-1}\cdot\frac{\Gamma(1/\kappa)\Gamma(-1/\kappa)}{\Gamma(N/\kappa)\Gamma(-N/\kappa)}$, where $\k\equiv N+k$. The four-point function of $g^{\a}_\b$ and its Hermitian conjugates is the solution of the well-known Knizhnik-Zamolodchikov equations \cite{Knizhnik:1984nr}
\[
&\la g^{\alpha}_{\beta}(z_1,\bar{z}_1)\left(g^\alpha_{\beta}(z_2,\bar{z}_2)\right)^\dagger g^{\alpha}_{\beta}(z_3,\bar{z}_3)\left(g^\alpha_{\beta}(z_4,\bar{z}_4)\right)^\dagger\ra_{\Sigma_1}\nn\\
=&\left\la g^{\alpha}_{\beta}(z_1,\bar{z}_1)(g^{-1})_\alpha^{\beta}(z_2,\bar{z}_2) g^{\alpha}_{\beta}(z_3,\bar{z}_3)(g^{-1})_\alpha^{\beta}(z_4,\bar{z}_4)\right\ra_{\Sigma_1}\nn\\
=&|z_{13}z_{24}|^{-4\Delta_g}\sum_{i,j,n=1,2}X_{nn}\mathcal{F}^{(n)}_i(\eta)\mathcal{F}^{(n)}_j(\bar{\eta}),
\]
where
\[
&X_{11}=1,\quad X_{22}=\frac{\Gamma \left(\frac{N+1}{\kappa }\right)\Gamma \left(\frac{N-1}{\kappa }\right) \Gamma \left(\frac{k}{\kappa }\right) \Gamma \left(\frac{k}{\kappa }\right) }{N^2 \cdot\Gamma \left(\frac{k+1}{\kappa }\right) \Gamma \left(\frac{k-1}{\kappa }\right) \Gamma \left(\frac{N}{\kappa }\right) \Gamma \left(\frac{N}{\kappa }\right)},\nn\\
&\mathcal{F}^{(1)}_1(\eta)=\eta^{-2\Delta_g}(1-\eta)^{\frac{N}{\kappa}-2\Delta_g}\cdot\mathbin{_2F_1}\big(1/\kappa,-1/\kappa,1-N/\kappa;\eta\big),\nn\\
&\mathcal{F}^{(1)}_2(\eta)=\frac{1}{k}\eta^{1-2\Delta_g}(1-\eta)^{\frac{N}{\kappa}-2\Delta_g}\cdot\mathbin{_2F_1}\big(1+1/\kappa,1-1/\kappa,2-N/\kappa;\eta\big),\nn\\
&\mathcal{F}^{(2)}_1(\eta)=\eta^{\frac{N}{\kappa}-2\Delta_g}(1-\eta)^{\frac{N}{\kappa}-2\Delta_g}\cdot\mathbin{_2F_1}\big((N-1)/\kappa,(N+1)/\kappa,1+N/\kappa;\eta\big),\nn\\
&\mathcal{F}^{(2)}_2(\eta)=-N\eta^{\frac{N}{\kappa}-2\Delta_g}(1-\eta)^{\frac{N}{\kappa}-2\Delta_g}\cdot\mathbin{_2F_1}\big((N-1)/\kappa,(N+1)/\kappa,N/\kappa;\eta\big).
\]
\begin{figure}[t]
	\centering
\captionsetup[subfloat]{farskip=5pt,captionskip=1pt}
\subfloat{
			\includegraphics[width =0.45\linewidth]{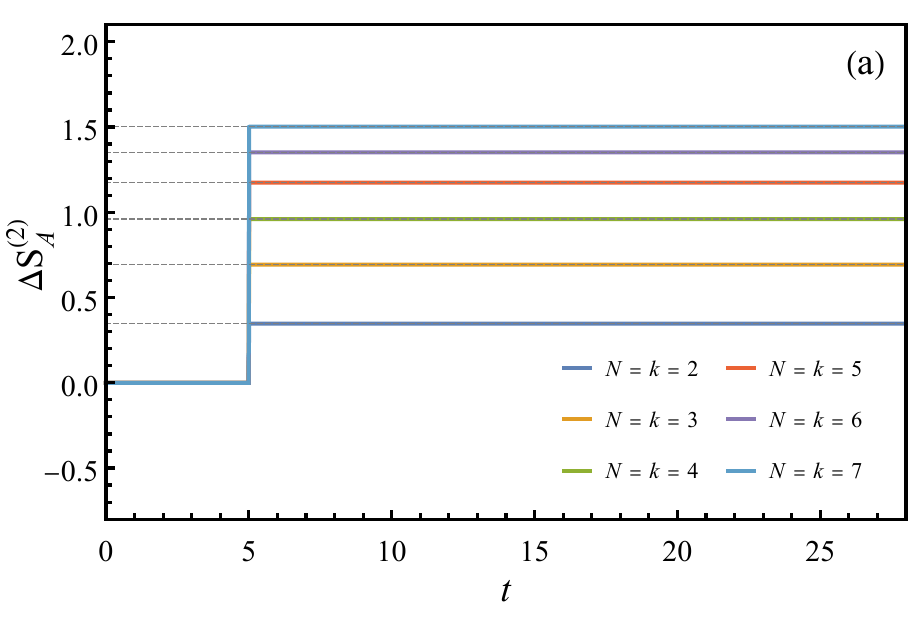}}
\hfill
\subfloat{
			\includegraphics[width =0.45\linewidth]{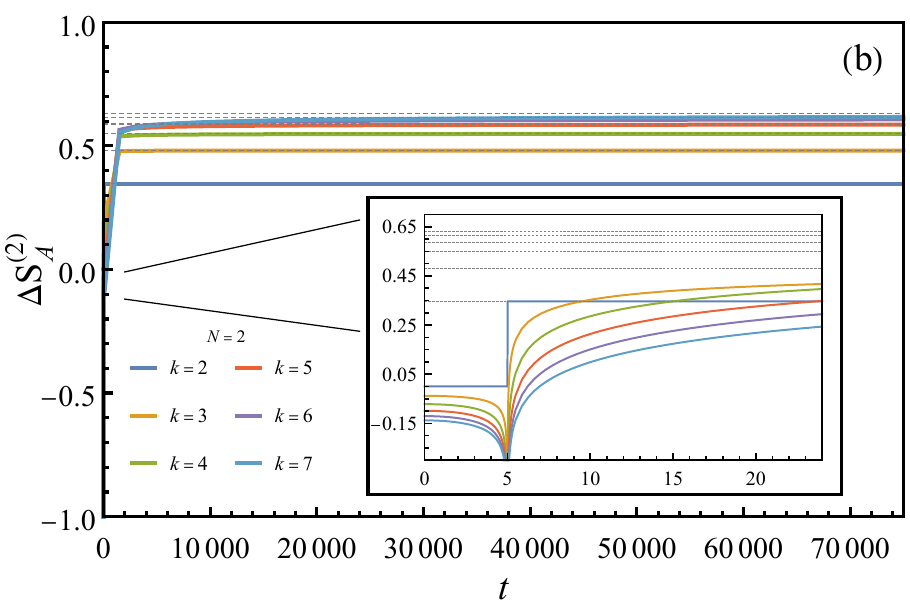}}\\

\subfloat{
			\includegraphics[width =0.45\linewidth]{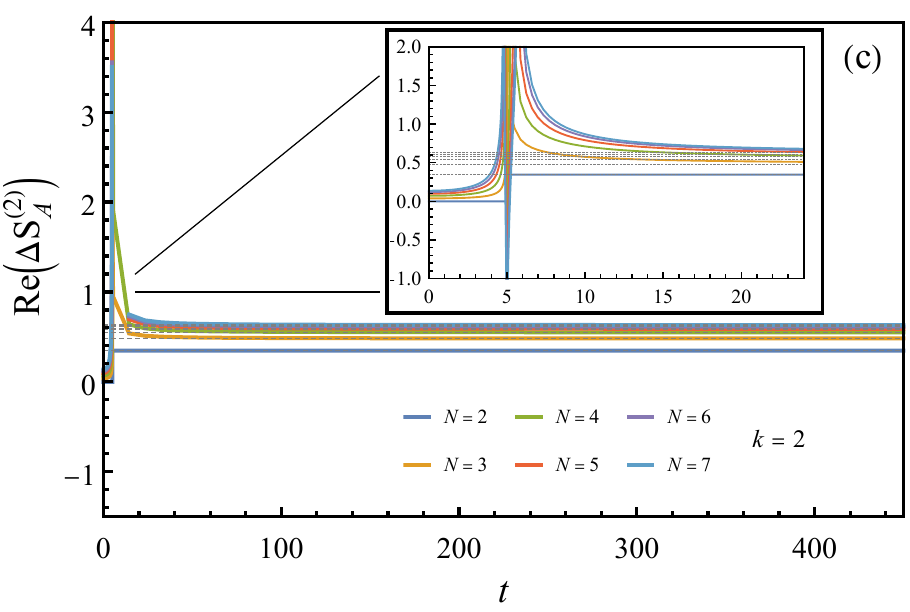}}
\hfill
\subfloat{
			\includegraphics[width =0.45\linewidth]{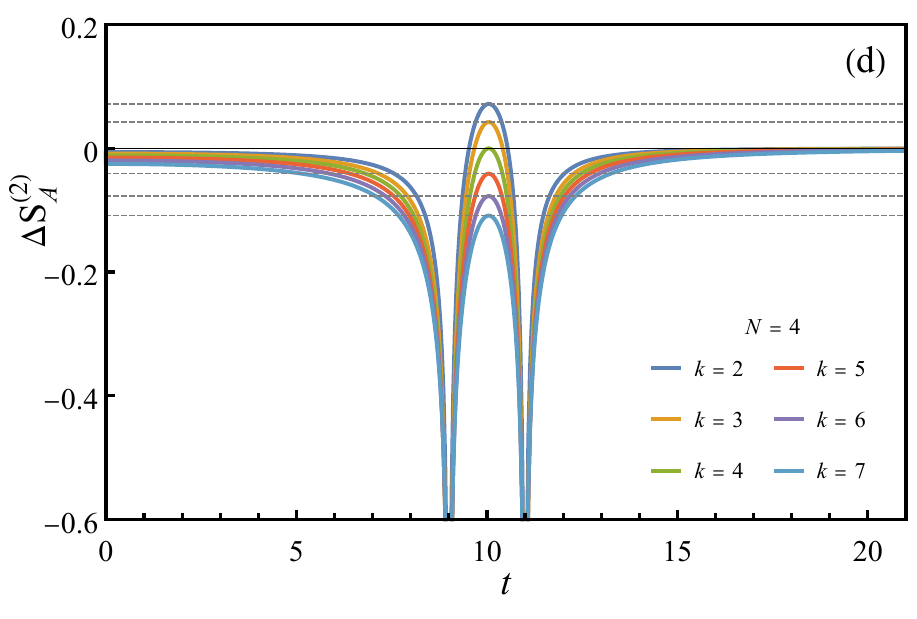}}
	\caption{The real-time evolution of $\Re[\Delta S_A^{(2)}]$ or $\Delta S_A^{(2)}$ in $\text{SU}(N)_k$ WZW models with $g^\a_\b$-excitation. The regulator is chosen to be $\e=10^{-5}$. (a), (b), (c): $A=[0,\infty)$, $l=10$, $x_m=0$. The dashed lines correspond to $\log d_g$ for different $N$ and $k$; (d): $A=[0,20]$, $x_m=10$, $l=2$. $\Delta S^{(2)}_A(\eta,\bar\eta)=\Delta S^{(2)}_A(1/2,1/2)$ at the dashed lines.}
	\label{Fig:WZW}
	\vspace{-0.5em}
\end{figure}
We next explore the full-time evolution behavior of $\Delta S_A^{(2)}$ utilizing the above information. Particularly, we are mainly concerned with the behavior of $\Delta S_A^{(2)}$ evolution under the two symmetric space configurations, i.e. $x_m=0$ for $A=[0,\infty)$ and $x_m=L/2$ for $A=[0,L]$. As shown in figure \ref{Fig:WZW}, we can see that in the case of $x_m=0$, $A=[0,\infty)$, there are three evolution patterns of $\Re[\Delta S^{(2)}_A]$ determined by the relative size of the rank $N$ and level $k$. When $N\leq k$ (see  figure \ref{Fig:WZW}(a) and (b)), we find that $\Tr[(\T^{1|2}_A(t))^2]$ is always greater than 0, thus $\Delta S_A^{(2)}$ remains real in the full-time evolution. A rather fascinating situation is that $N=k$, since again we observe a pseudo-R\'enyi entropy evolution identical to that of R\'enyi entropy. When  $N>k$, the $2$nd pseudo-R\'enyi entropy evolution behavior is similar to that in the minimal model. Since $\Tr[(\T^{1|2}_A(t))^2]$ is less than 0 near $t=l/2$, we have a complex pseudo entropy of in a certain interval. Figure \ref{Fig:WZW}(d) depicts the behaviors of $S_A^{(2)}$ under the second symmetric space configuration $x_m=L/2$ in $A=[0,L]$. Notice that no matter $k$ is greater than $N$ or not,  we obtain a pseudo-R\'enyi entropy that remains real throughout time evolution.
\paragraph{ summaries of the above results:}
Let us take a short stay to briefly summarize the above results and try to answer the questions posed at the beginning of the section.
\begin{enumerate}
 \item In general, there are one or two hump evolutions (for example, figure \ref{Fig:FB}(a), (b)) between the early time and late time evolution  of $\Delta S_A^{(2)}$, and the value of $\Delta S_A^{(2)}$ at the peaks of humps is $\Delta S_A^{(2)}(\eta,\bar\eta)=\Delta S_A^{(2)}(\frac{1}{2},0)$.
  \item We find two special spatial configurations of operators, $x_1=-x_2$ for $A=[0,\infty)$ and $x_1=L-x_2$ for $A=[0,L]$, in which the  the trace of $\big(\T^{1|2}_A(t)\big)^2$ always remains real. Within them, for the case of $x_1=L-x_2$, we expect $\Tr\big[\big(\T^{1|2}_A(t)\big)^2\big]$ to be greater than 0, resulting in a real $\Delta S_A^{(2)}$, which is consistent with all the numerical results.
  \item In the $\text{e}^{\frac{i}{2}\phi}$+$\text{e}^{-\frac{i}{2}\phi}$-excitation of free scalar and $g_\b^\a$-excitation of $\text{SU}(N)_k$ WZW models ($N=k$), we observe  that the $2$nd pseudo-R\'enyi entropy exhibits the same behavior as R\'enyi entropy in the case of $x_1=-x_2$, $A=[0,\infty)$, i.e.
  \[
\Delta S_{[0,\infty)}^{(2)}(x_1,-x_1,t)=
\begin{cases}
0,&0\leq t<|x_1|,\\
\log d,&t>|x_1|.\label{midconclusion}
\end{cases}
\]
\end{enumerate}
\subsection{Linear combination of operators}
In the previous subsection, we study the real-time behaviors of $\Delta S_A^{(2)}$ for states excited by the same primary operator. It's not so straightforward to extend the results
to two different primary operators. Simply substituting one of the operators would probably make a non-normalizable transition matrix since the two-point function for two different primary operators is likely to be zero.
One feasible way is to consider the linear combinations of operators.\footnote{We thank Tadashi Takayanagi for bringing this idea to our attention.} Consider a real-time dependent transition matrix $\mathcal{T}^{\psi|\tilde \psi}(t)=\frac{\text{e}^{-iHt}|\psi\rangle\langle\tilde\psi|e^{iHt}}{\langle\tilde\psi|\psi\rangle}$ consisting of two quantum states $|\psi\rangle$ and $|\tilde\psi\rangle$,
\begin{align}
&|\psi\rangle:=\frac{1}{\sqrt{\la\mo^\dagger(x,\e)\mo(x,-\e)\ra}}\mo(x,-\e)|\Omega\rangle,
\quad|\tilde\psi\rangle:=\frac{1}{\sqrt{\la\tilde\mo^\dagger(\tilde x,\e)\tilde\mo(\tilde x,-\e)\ra}}\tilde\mo(\tilde x,-\e)|\Omega\rangle,\nn\\
&\mo(x,-\e)=\sum_pC_p\mo_p(x,-\e),~~~~~~~~~~~~~~~~~~~\tilde\mo(\tilde x,-\e)=\sum_p\tilde C_p\mo_p(\tilde x,-\e).
\end{align}
In the above, $\mo(x,\e)\equiv\text{e}^{\e H}\mo(x)\text{e}^{-\e H}$, $\mathcal{O}_p$ are primary or descedant operators that are orthogonal to each other in the sense of the two-point function, $C_p$($\tilde C_p$) are superposition coefficients used to give a non-zero inner product.
\subsubsection{The expected late time limit of $\Delta S^{(n)}_A$}
 Let's focus on the case where $A=[0,\infty)$, we expect the late time limit of $\Delta S_A^{(n)}$ to take the following form
\[
\lim_{t\to\infty}\Delta S^{(n)}[\mathcal{T}_A^{\psi|\tilde\psi}(t)]=\frac{1}{1-n}
\log\left[\sum_p\left(\frac{C_p\tilde C_p^*\langle\mathcal{O}^\dagger_p(\tilde w,\bar {\tilde{w}})
\mathcal{O}_p(w,\bar {{w}})\rangle}{\sum_{p'}C_{p'}\tilde C^*_{p'}\langle\mathcal{O}^\dagger_{p'}(\tilde w,\bar{\tilde{w}})
\mathcal{O}_{p'}(w,\bar w)\rangle}\right)^n\text{e}^{(1-n)S^{(n)}[\mathcal{O}_p]}\right],\label{resultsfinal}
\]
where $w=x-i\epsilon$, $\tilde w=\tilde x+i\epsilon$, and $S^{(n)}[\mathcal{O}_p]$ is the late time limit
of the difference of entanglement entropy of $\mathcal{O}_p$-excitation ($S^{(n)}[\mathcal{O}_p]=\log d_p$ in RCFTs). It is difficult to utilize replica trick to prove Eq.\eqref{resultsfinal}. Nevertheless, we can provide a quantum mechanical derivation from another perspective, which as far as we know was first introduced in \cite{Guo:2018lqq}.\footnote{The derivation is presented in appendix \ref{app2:derivation}.} We next numerically examine the correctness of Eq.\eqref{resultsfinal} using the replica trick in the concrete model.

\subsubsection{Example in critical Ising}
We would like to compute $\Delta S_A^{(2)}$ of linear combination operators in the critical Ising model to examine Eq.\eqref{resultsfinal}. There are three primary operators in the Ising model at a critical point, namely the identity $\mathbb{I}$, the spin $\sigma$, and the energy density $\ve$. The fusion rule of them is well-known,
\[
\ve\times\ve=\mathbb{I},\quad\s\times\s=\mathbb{I}+\ve,\quad \s\times\ve=\s.\label{fusionrule}
\]
For simplicity, below, we consider the combination of $\s$ and $\mathbb{I}$ as a typical example.
\paragraph{Example---$\sigma$+$\mathbb{I}$:}
Let us first define two linear combination operators
\[
\mo(w,\bar w)\equiv C_\s\cdot\sigma(w,\bar w)+C_{\Id}\cdot\Id,\quad \tilde\mo(\tilde w,\bar{\tilde{w}})\equiv\tC_\s\cdot\sigma(\tilde w,\bar{\tilde{w}})+\tC_\Id\cdot\mathbb{I}.
\]
According to the fusion rule and \eqref{2thPEE}, only four- and two-point functions of $\s$ are involved in the calculation,
\[
\la\s(z_1,\bar z_1)\s(z_2,\bar z_2)\ra_{\Sigma_1}=&\frac{1}{|z_{12}|^{1/4}},\nn\\
\la\s(z_1,\bar z_1)\s(z_2,\bar z_2)\s(z_3,\bar z_3)\s(z_4,\bar z_4)\ra_{\Sigma_1}=&\Bigg(\frac{1}{2}\left|\sqrt{\frac{z_{14}z_{23}}{z_{12}z_{34}z_{13}z_{24}}}\right|+\frac{1}{2}\left|\sqrt{\frac{z_{13}z_{24}}{z_{12}z_{34}z_{14}z_{23}}}\right|\nonumber\\
&+\frac{1}{2}\left|\sqrt{\frac{z_{12}z_{34}}{z_{13}z_{24}z_{14}z_{23}}}\right|\Bigg)^\frac{1}{2}.
\]
In addition,  since in general $\Delta S_A^{(2)}$ of the mixed operator cannot be written as a function of cross ratios, we choose to use the following coordinates mapping between $(w,\bar w)$ on $\Sigma_2$ and $(z,\bar z)$ on $\Sigma_1$\cite{He:2014mwa} to complete the analytic continuation of time,
\[
z_1=\sqrt{w_1}=i\sqrt{-x_1-t+i\e},\quad \bar z_1=\sqrt{\bar w_1}=-i\sqrt{-x_1+t-i\e},\\
z_2=\sqrt{w_2}=i\sqrt{-x_2-t-i\e},\quad \bar z_2=\sqrt{\bar w_2}=-i\sqrt{-x_2+t+i\e}.
\]

\begin{figure}[h]
	\centering
\captionsetup[subfloat]{farskip=5pt,captionskip=1pt}
\subfloat{
			\includegraphics[width =0.45\linewidth]{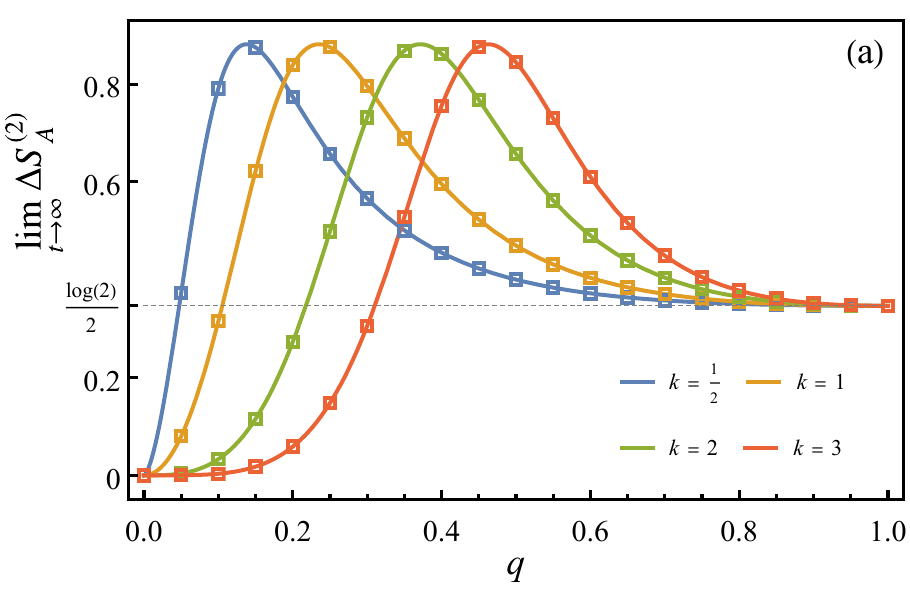}}
\hfill
\subfloat{
			\includegraphics[width =0.45\linewidth]{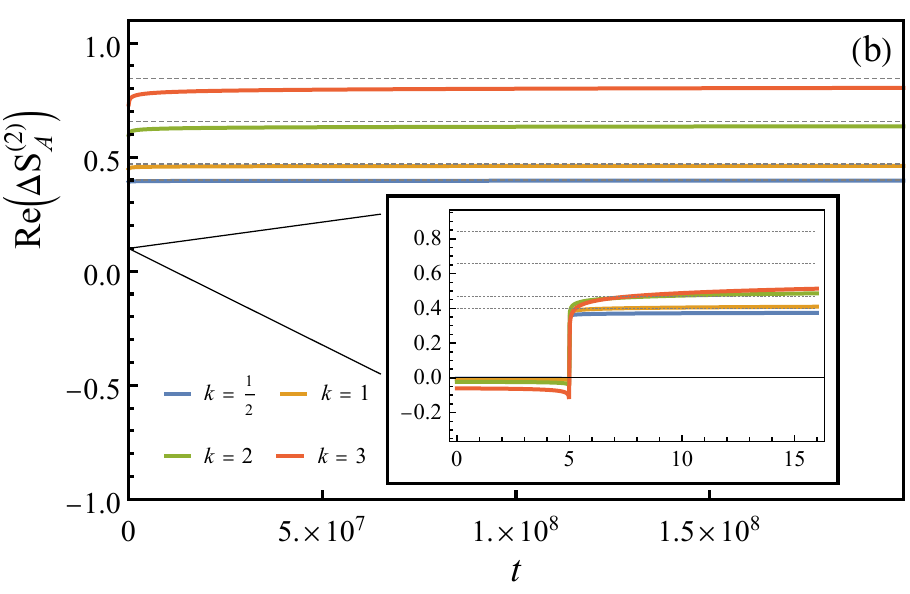}}\\

\subfloat{
			\includegraphics[width =0.45\linewidth]{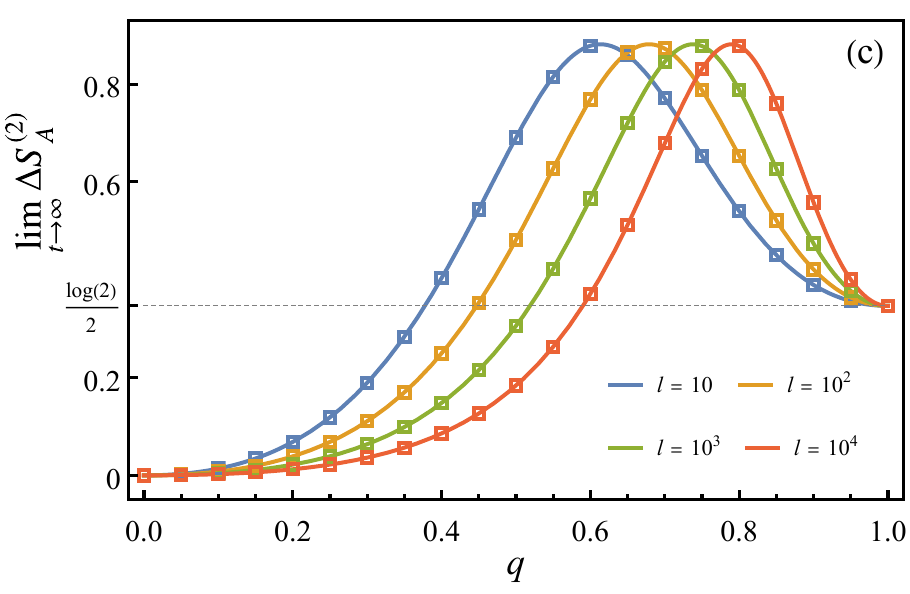}}
\hfill
\subfloat{
			\includegraphics[width =0.45\linewidth]{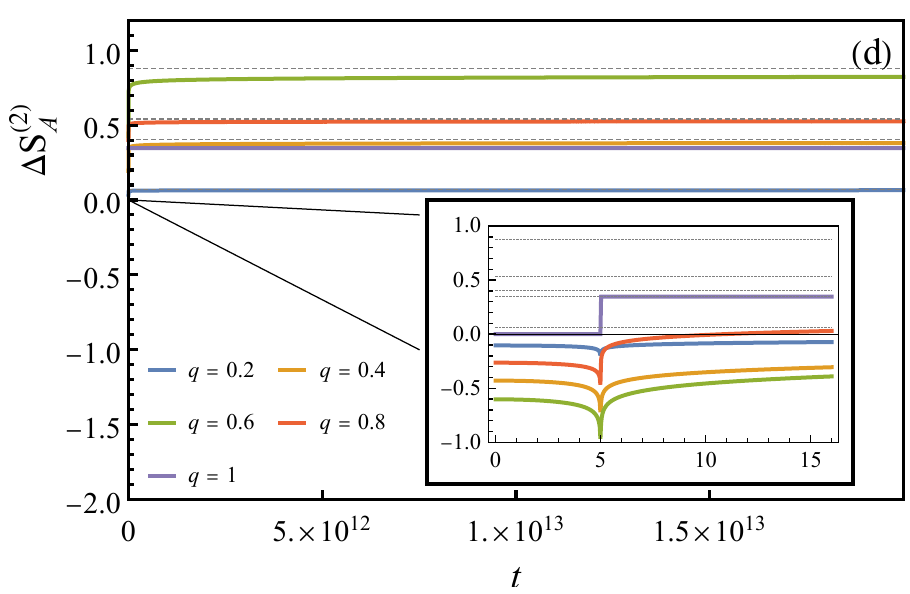}}
	\caption{(a) and (c): The late time limits of $\Delta S_A^{(2)}$ with respect to the mixing factor $q$ in $\sigma$+$\Id$-excitation. The regulator is chosen to be  $\e=10^{-5}$;~(b): The full-time evolution of $\Re[\Delta S_A^{(2)}]$ in $\s$+$\Id$-excitation. Parameters are selected as $q=0.5$, $x=\tilde x=-5$, $\e=10^{-5}$. The dashed lines are the theoretical limits derived from Eq.\eqref{resultsfinal} for the corresponding parameters;~(d): The full-time evolution of $\Delta S_A^{(2)}$ in $\s$+$\Id$-excitation. Parameters are selected as $x=-\tilde x=\pm5$, $\e=10^{-5}$. The dashed lines are the theoretical limits derived from Eq.\eqref{resultsfinal} for the corresponding parameters.}
	\label{Fig:sigma+I}
	\vspace{-0.5em}
\end{figure}

We start with the case of $\{\tilde C_p\}\neq\{C_p\}$ and $\tilde x=x$. An efficient way is to set $C_{\s}=q\in[0,1]$, $C_\Id=1-q$, $\tC_\s=q^k$, $\tC_{\Id}=1-q^k$, and obviously what we will obtain when $k\neq1$ is pseudo-R\'enyi entropy rather than R\'enyi entropy. Figure \ref{Fig:sigma+I}(a) shows the behavior of the late time limits of $\Delta S_A^{(2)}$ when we adjust the mixed coefficient $q$. We can see that the late time limits of $\Delta S_A^{(2)}$ obtained by replica method numerically (square points) are in good agreement with the results (solid lines) given by Eq.\eqref{resultsfinal}. With the increase of $q$, the contribution of $\s$ operator gradually increases, which leads to the saturation value of 2nd (pseudo-) R\'enyi entropy  gradually shifting from $0~(\log d_{\mathbb{I}})$ to $\log\sqrt{2}~(\log d_{\s})$. Except for the late time limit, it's also intriguing to depict the the full-time evolution of $\Delta S^{(2)}_A$, see figure \ref{Fig:sigma+I}(b). We find that although $\Delta S_A^{(2)}$ saturates to a real value, globally $\Delta S_A^{(2)}$ is complex in all cases except $k=1$ (the case of R\'enyi entropy).

The other interesting case we shall investigate is that $\{\tilde C_p\}=\{C_p\},~\tilde x\neq x$. Let us set $\tilde C_{\s}=C_{\s}=q$, $\tilde C_{\Id}=C_{\Id}=1-q$. According to \eqref{resultsfinal}, since $\langle\mo_p^\dagger(\tilde w,\bar{\tilde w})\mo_p(w,\bar w)\rangle=\big((\tilde x-x)^2+4\e^2\big)^{-2\Delta_p}$, only the information of the distance of two space points $l\equiv|\tilde x -x|$ is related. We then plot the change of saturation value of $\Delta S^{(2)}_A$ with $q$ under different $l$, as depicted in figure \ref{Fig:sigma+I}(c). Once again, we find that the theoretical value (solid lines) given by Eq.$\eqref{resultsfinal}$ is consistent with the numerical result (square points) given by replica trick. On the other hand, when the insertion point is symmetric about the origin, i.e. $\tilde x=-x$, $\Delta S_A^{(2)}$ is found to be real throughout the  the time evolution (see figure \ref{Fig:sigma+I}(d)), which is consistent with the previous symmetry analysis.
\section{General arguments and examples on $\Delta S^{(n)}_{A}$}\label{sec4}
\begin{figure}[t]
  \centering
  \includegraphics[width =0.8\linewidth]{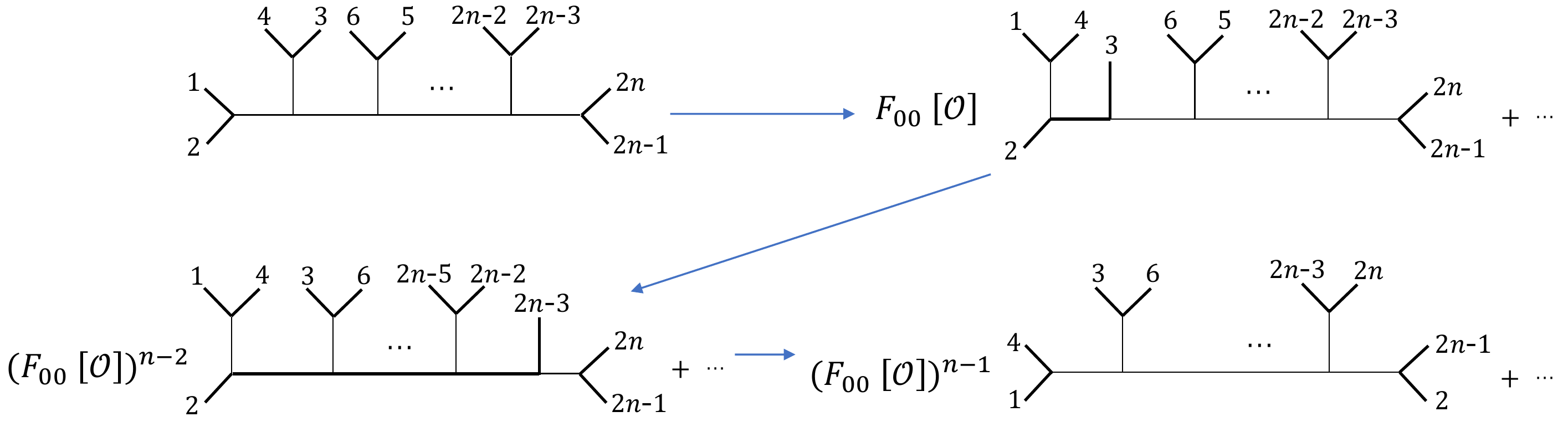}
   \caption{The fusion transformations to obtain $\Delta S_A^{(n)}$.}\label{Fig:Fmatrix}
\end{figure}

In the previous section, we study the real-time behavior of $\Delta S_A^{(2)}$ for two insertion operators with different spatial coordinates. Meanwhile, we propose a formula to describe the late time limit of $\Delta S_A^{(n)}$ of linear combination operators. However, as shown in appendix \ref{app2:derivation},  its rationality also depends on the behavior of $n$th pseudo-R\'enyi entropy of a single primary operator
insertion. Therefore, in this section, we shall quest for the properties of $\Delta S_A^{(n)}$ for two operators with different space points in the light of the results of $\Delta S_A^{(2)}$ that we have found before.
\paragraph{Late time limit of $\Delta S_A^{(n)}$:}
The excess of $n$th pseudo-R\'enyi entropy of the reduced transition matrix $\T_A^{1|2}(t)$ \eqref{TforsingleO} can be obtain from Eq.\eqref{mthPEEinpathintegtrallanguage} by computing the $2n$-point function on the replica manifold $\Sigma_n$. Our primary purpose is to explore the existence of the late time saturation value of $\log d_\mo$ in pseudo-R\'enyi entropy of higher-order when the subsystem $A=[0,\infty)$. Hence we employ the conformal map $\eqref{map2}$ to obtain the coordinates of $2n$ operators on $\Sigma_1$ first,
\[
z_{2k+1}=&\text{e}^{2\pi i\frac{k+1/2}{n}}(-x_1-t+i\e)^{\frac{1}{n}},\quad \bar z_{2k+1}=\text{e}^{-2\pi i\frac{k+1/2}{n}}(-x_1+t-i\e)^{\frac{1}{n}},\nn\\
z_{2k+2}=&\text{e}^{2\pi i\frac{k+1/2}{n}}(-x_2-t-i\e)^{\frac{1}{n}},\quad \bar z_{2k+2}=\text{e}^{-2\pi i\frac{k+1/2}{n}}(-x_2+t+i\e)^{\frac{1}{n}},\quad (k=0,...,n-1).\label{A=INFzcoordin}
\]
Then we have
\[
\frac{\langle\mo^\dagger(w_{2n},\bar w_{2n})\mo(w_{2n-1},\bar w_{2n-1})...\mo(w_1,\bar w_1)\rangle_{\Sigma_n}}{\langle\mo^\dagger(w_2,\bar w_2)\mo(w_1,\bar w_1)\rangle^n_{\Sigma_1}}=\mathcal{C}_n\cdot\langle\mo^{\dagger}(z_{2n},\bar z_{2n})\mo(z_{2n-1},\bar z_{2n-1})...\mo(z_1,\bar z_1)\rangle_{\Sigma_1},\label{logdon}
\]
where
\[
\mathcal{C}_n\equiv&\left(\frac{(x_1-x_2)^2+4\e^2}{n^2}\right)^{2n\Delta_\mo}\times\prod^{2n}_{i=1}(z_i^{n-1}\bar z_i^{n-1})^{-\Delta_\mo}\nn\\
\simeq&\left(\frac{(x_2-x_1)^2+4\e^2}{n^2}\right)^{2n\Delta_\mo}\times t^{4(1-n)\Delta_\mo}+\text{sub-leading order terms}\label{cninooo}\\
\simeq&0\quad(t\to\infty).\nn
\]
On the other hand, we find that at the late time ($t\to\infty$)
\[
&\lim_{t\to\infty}(z_{2(k+1)+2}-z_{2k+1})\simeq\frac{x_2-x_1+2i\e}{nt}e^{2\pi i\frac{k+1}{n}}t^{\frac{1}{n}}\simeq0,\nn\\
&\lim_{t\to\infty}(\bar z_{2k+2}-\bar z_{2k+1})\simeq\frac{x_1-x_2+2i\e}{nt}e^{-2\pi i\frac{k+1/2}{n}}t^{\frac{1}{n}}\simeq0,\quad(k=0,...,n-1;~z_{2n+2}\equiv z_{2n}).
\]
The above results enable us  to factorize the $2n$-point function $\la\mo^\dagger(z_{2n},\bar z_{2n})...\mo(z_{1},\bar z_{1})\ra_{\Sigma_1}$ into $n$-point functions by using the fusion transformation \eqref{fusionransformation} $n-1$ times (see figure \ref{Fig:Fmatrix}),
\[
&\la\mo^\dagger(z_{2n},\bar z_{2n})...\mo(z_{1},\bar z_{1})\ra_{\Sigma_1}\nn\\
\simeq&(F_{00}[\mo])^{n-1}\times\left(\prod_{k=0}^{n-1}(z_{2k+4}-z_{2k+1})(\bar z_{2k+2}-\bar z_{2k+1})\right)^{-2\Delta_\mo}+\text{sub-leading order terms}\nn\\
\simeq&(F_{00}[\mo])^{n-1}\times\left(\frac{(x_2-x_1)^2+4\e^2}{n^2}\right)^{-2n\Delta_\mo}\times t^{4(n-1)\Delta_\mo}+\text{sub-leading order terms}
\label{onsigma1}
\]
Substituting \eqref{cninooo} and \eqref{onsigma1} into the r.h.s. of Eq.\eqref{logdon}, it's easy to find that the leading-order contribution at late time is $(F_{00}[\mo])^{n-1}=d_{\mo}^{1-n}$. In this way, we obtain  the late time value of $\Delta S_A^{(n)}$
\[
\lim_{t\to\infty}\Delta S_A^{(n)}=\log d_\mo.
\]

\begin{figure}[htpb]
	\centering
\captionsetup[subfloat]{farskip=5pt,captionskip=1pt}
\subfloat{
			\includegraphics[width =0.45\linewidth]{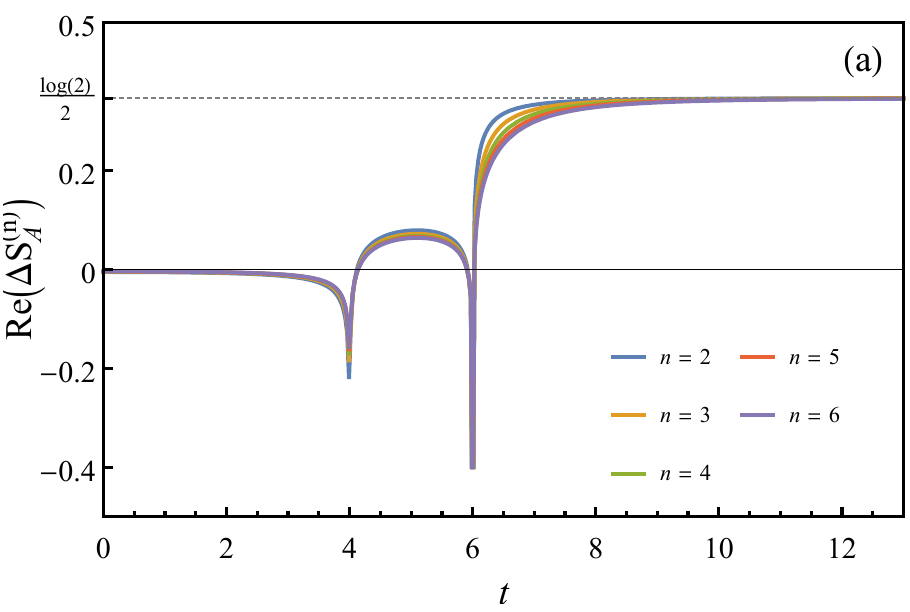}}
\hfill
\subfloat{
			\includegraphics[width =0.45\linewidth]{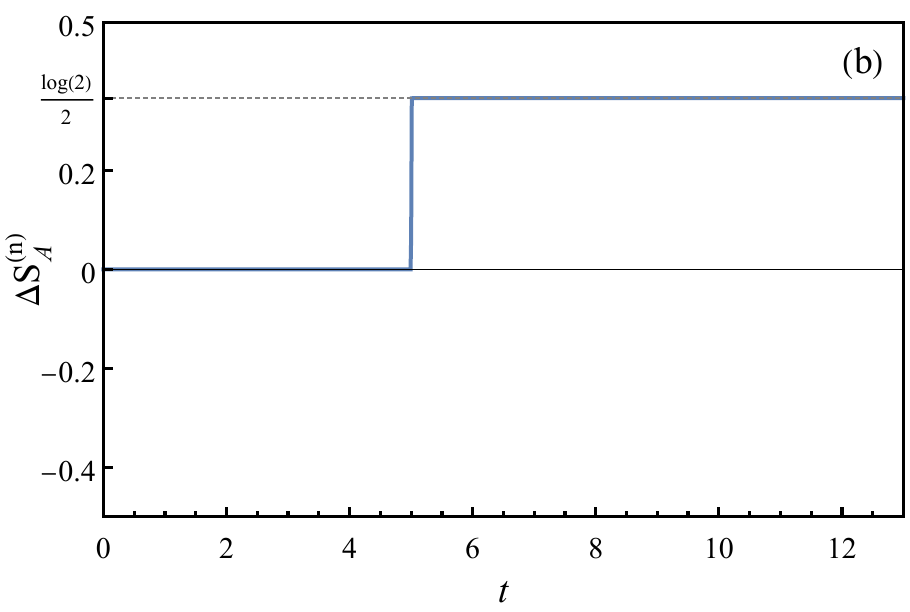}}\\

\subfloat{
			\includegraphics[width =0.45\linewidth]{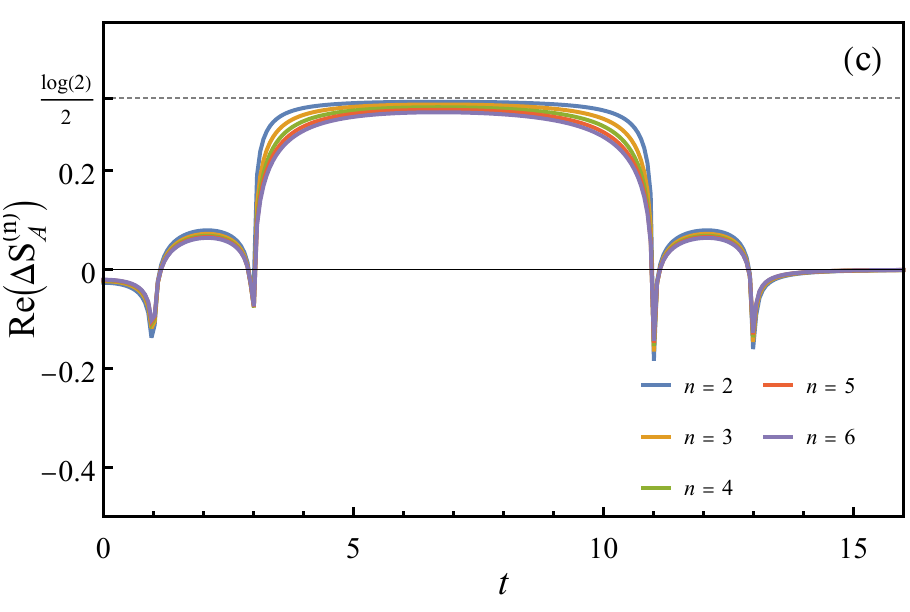}}
\hfill
\subfloat{
			\includegraphics[width =0.45\linewidth]{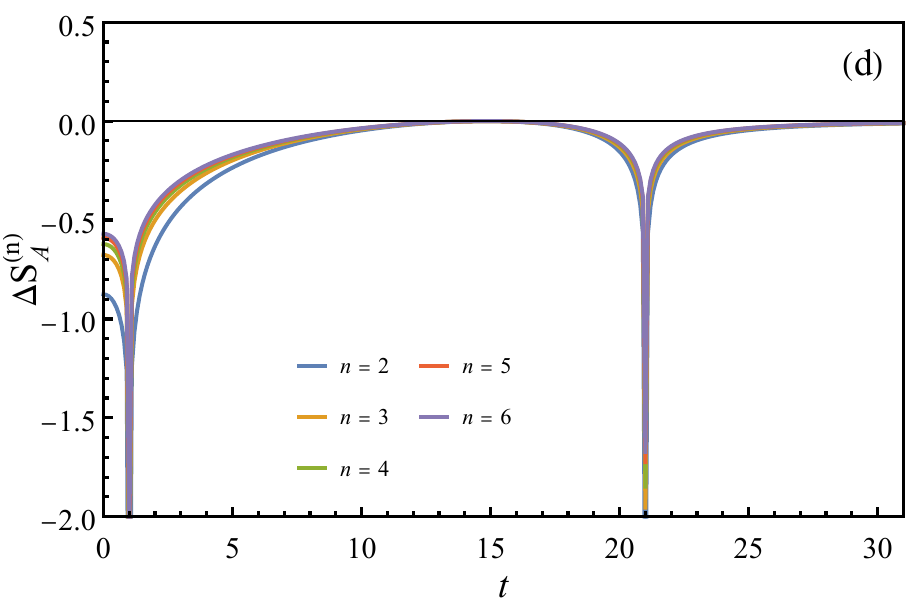}}
	\caption{The real-time evolution of $\Re[\Delta S_A^{(n)}]$ or $\Delta S_A^{(n)}$ for $\sigma$-excitation in critical Ising model. The regulator $\e=10^{-5}$. (a): $A=[0,\infty)$, $x_m=\pm5$, $l=2$; (b): $A=[0,\infty)$, $x_m=0$, $l=10$; (c): $A=[0,10]$, $x_m=-2$ or $12$, $l=2$; (d): $A=[0,20]$, $x_m=10$, $l=22$.}
	\label{Fig:nthSA}
	\vspace{-0.5em}
\end{figure}

\paragraph{Symmetries of $\Delta S^{(n)}_A$:} The second intention in this section is to investigate whether the symmetries found in second pseudo-R\'enyi entropy, i.e. Eq.(\ref{sym1forSA}-\ref{SA=INF}), still hold in higher-order or not. It may be difficult to verify analytically, but the numerical examination is easy to take. One good object of study is the $\s$-excitation in the critical Ising  model, since the $2n$-point function of the spin operator $\s$ is well-known \cite{Luther:1975wr,DiFrancesco:1987ez},
\[
\langle\s(z_1,\bar z_1)...\s(z_{2n}, \bar z_{2n})\rangle_{\Sigma_1}=\frac{1}{2^{n}}\sum_{\ve_i=\pm1~i=1,...,2n\atop \sum\ve_i=0}~\prod_{i<j}|z_i-z_j|^{\ve_i\ve_j/2}.\label{nptsigma}
\]
With the help of \eqref{A=INFzcoordin} and \eqref{nptsigma}, we can study the evolution behavior of $\Delta S_A^{(n)}$ under the first symmetric space configuration, i.e. $x_1=-x_2$,  $A=[0,\infty)$.
On the other hand, for the second symmetric space configuration of $x_1=L-x_2$ in $A=[0, L]$,  the coordinates of 2n operators on $\Sigma_1$ will change to the following form
\[
z_{2k+1}=&e^{\frac{2\pi i k}{n}}\left(\frac{x_1+t-i\e}{x_1+t-i\e-L}\right)^{\frac{1}{n}},\quad\bar z    _{2k+1}=e^{-\frac{2\pi i k}{n}}\left(\frac{x_1-t+i\e}{x_1-t+i\e-L}\right)^{\frac{1}{n}},\nn\\
z_{2k+2}=&e^{\frac{2\pi i k}{n}}\left(\frac{x_2+t+i\e}{x_2+t+i\e-L}\right)^{\frac{1}{n}},\quad\bar z    _{2k+2}=e^{-\frac{2\pi i k}{n}}\left(\frac{x_2-t-i\e}{x_2-t-i\e-L}\right)^{\frac{1}{n}},\quad (k=0,...,n-1).
\]
Figure \ref{Fig:nthSA} demonstrates all situations that we are interested in. We find that the symmetries (\ref{sym1forSA}-\ref{SA=INF}) also hold in the higher-order pseudo-R\'enyi entropy. It can be clearly seen from figure \ref{Fig:nthSA}(b) and (d). Because we know that the establishment of(\ref{sym1forSA}-\ref{SA=INF}) may bring about a real $\Delta S^{(n)}_A$ evolution. Another interesting finding  is that (b) shows that the evolution of higher-order pseudo-R\'enyi entropy of  $\s$-excitation under the first symmetric space configuration  still maintains the evolution pattern described by \eqref{midconclusion}.\footnote{Due to the increasing computational complexity, we verify this point up to $n=7$.} We can also explore the  asymmetric cases, as shown in (a) and (c), and there are two
points worth noting: i). The higher-order pseudo-R\'enyi entropy in asymmetric space configuration still has hump evolution, and its peak value changes with $n$; ii). Figure \ref{Fig:nthSA}(c) suggests that after the relative sizes of $L$ and $l$ are fixed, the middle time behavior of $\log d$ will gradually disappear with the increase of $n$.

\section{Conclusions and prospect}\label{sec5}
In this work, we study a generalized version of entanglement entropy and R\'enyi entropy, which are so-called pseudo-entropy (PE)  and pseudo-R\'enyi entropy (PRE), respectively, in 2d CFTs. In particular, the real-time evolution of PRE associated with two locally excited states has been evaluated in various 2d CFTs, e.g., free bosonic field theory, critical Ising model, WZW model as well as large-$c$ CFTs. These locally excited states are generated by acting local operators on the vacuum state, and these operators can be a single primary operator or a linear combination of them.
Some fascinating behaviors of PRE evolution are found as follows:

For the reduced transition matrix generated by two primary operators with different spatial coordinates \eqref{TforsingleO}, we show that when subsystem $A$ has an infinite length, the late time value of $2$nd PRE is logarithmically divergent in large-$c$ CFTs (take the large $c$ limit first). The late time value of $n$th PRE saturates to $\log d$ in RCFTs (for Example, see figure \ref{Fig:nthSA}(a)), where $d$ is the quantum dimension of the corresponding primary operator. Whereas, when subsystem $A$ has a finite length, we show that the middle time behavior of $\log d$ of PRE in RCFTs gradually disappears as the distance between operators or the order number $n$ increases (see figure\ref{Fig:FB}(c) and figure \ref{Fig:nthSA}(c) respectively).
Unlike the entanglement entropy, we find that in general, there is a hump during the evolution between the early and late time evolution of the $n$th PRE (for example, see figure \eqref{Fig:nthSA}(a)), and for $n=2$, its peak value can be found as $\Delta S_A^{(2)}(\eta,\bar\eta)=\Delta S_A^{(2)}(1/2,0)$, where $(\eta,\bar\eta)$ are cross ratios.

 On the other hand,  for excitation by a linear combination of operators, using Schmidt decomposition, we find that the late time limit of the $n$th pseudo-R\'enyi entropy is governed by the formula \eqref{resultsfinal}. A prominent property that distinguishes linear combination excitation from single primary operator excitation is that the late limit of PRE under linear combination excitation is not necessarily the same as that of R\'enyi entropy (see, for example, figure \ref{Fig:sigma+I}(c)). This means that for the case of a single primary operator, the initial information about the positions of the insertion operators is lost in the long-time evolution. In contrast, for the case of linear combinations, the late limit of the pseudo-R\'enyi entropy still contains the initial information of the operator positions. It would be interesting to explore whether it is possible to recover the initial data by using the late time limit of pseudo-R\'enyi entropy.

Finally, building on the analysis of the cross ratios,  we uncover three kinds of symmetries for the $2$nd PRE (\ref{sym1forSA}-\ref{SA=INF}), which naturally screen out two kinds of special space configurations of insertion operators---$x_1=-x_2$ for subsystem $A=[0,\infty)$ and $x_1=L-x_2$ for subsystem $A=[0,L]$. We show that the trace of $\big(\T^{(2)}_A(t)\big)^2$ is always real in both configurations,  and we expect $\Tr\big[\big(\T^{(2)}_A(t)\big)^2\big]$ to be positive in the second configuration, which gives us a real $2$nd PRE evolution.  For the first configuration, although the evolution of PRE under it is not always real, in some theories, the evolution of PRE under the first configuration in $A=[0,\infty)$ shows the same evolutionary pattern as R\'enyi entropy (see figure \ref{Fig:FB}(a) in the free scalar theory, figure \ref{Fig:WZW}(a) in the WZW models, and figure \ref{Fig:nthSA}(b) in the critical Ising model), i.e., we may have
\[
\Delta S_{A=[0,\infty)}^{(n)}(x,-x,t)=
\begin{cases}
0,&0\leq t<|x|,\\
\log d,&t>|x|.
\end{cases}
\]

It will be an attractive research direction for us in the future to fully clarify the condition that PRE remains real in the time evolution process and the condition that PRE behaves as R\'enyi entropy in the first symmetric space configuration. Furthermore, it's also interesting to make a  higher-dimensional generalization of our results and dig out the possible corresponding holographic counterpart.

\subsection*{Acknowledgements}
We thank Tadashi Takayanagi and Yuan Sun for valuable conversations and correspondence. SH would like to appreciate the financial support from Jilin University, Max Planck Partner group, and Natural Science Foundation of China Grants (No.12075101, No.1204756). WZG is supposed by the National Natural Science Foundation of China under
Grant No.12005070 and the Fundamental Research Funds for the Central
Universities under Grants NO.2020kfyXJJS041.
\appendix
\section{Several time limits of cross ratios}
In this appendix, we will analyze the cross ratios in various limits under two configurations of the subsystem $A$.
\subsection{$A=[0,\infty)$}\label{appen-a1}
For the case of $A=[0,\infty)$, the cross ratios in Euclidean signature  can be expressed in polar coordinates on $\Sigma_2$ as follows
\[
\eta=&\frac{1}{2}-\frac{(r_1+r_2)\cos\big(\frac{\theta_1-\theta_2}{2}\big)+i(r_1-r_2)\sin\big(\frac{\theta_1-\theta_2}{2}\big)}{4\sqrt{r_1}\sqrt{r_2}},\nn\\
\bar{\eta}=&\frac{1}{2}-\frac{(r_1+r_2)\cos\big(\frac{\theta_1-\theta_2}{2}\big)-i(r_1-r_2)\sin\big(\frac{\theta_1-\theta_2}{2}\big)}{4\sqrt{r_1}\sqrt{r_2}},\label{appA1,etabeforeac}
\]
where  $(r_1\cos\theta_1,r_{1}\sin\theta_1)=(x_1,-\t_{1}),~(r_2\cos\theta_2,r_{2}\sin\theta_2)=(x_2,\t_{2}),~(0\leq\theta_{j}<2\pi,~j=1,2)$, and
\[
\cos(\theta_1-\theta_2)=2\cos^2\left(\frac{\theta_1-\theta_2}{2}\right)-1=1-2\sin^2\left(\frac{\theta_1-\theta_2}{2}\right)=\frac{x_1x_2-\t_1\t_2}{\sqrt{(x_1^2+\t_1^2)}\sqrt{(x_2^2+\t_2^2)}}.
\]
Since the two Euclidean times $\t_1$ and $\t_2$ after analytic continuation are $\e+it$ and $\e-it$, respectively, we may set $\t_1=\t_2=\e$ before the analytic continuation. It leads to two ranges of $\theta_1-\theta_2$ which depend on $x_1+x_2$: $0<\theta_1-\theta_2<\pi$ when $x_1+x_2<0$; $\pi<\theta_1-\theta_2<2\pi$ when $x_1+x_2>0$. Thus we have
the following expressions of $\sin(\cos)\left(\frac{\theta_1-\theta_2}{2}\right)$ after the analytic continuation
\[
\cos\left(\frac{\theta_1-\theta_2}{2}\right)&=\left(\frac{1}{2}+\frac{x_1x_2-\e^2-t^2}{2\sqrt{(x_1^2+(\e+it)^2)}\sqrt{(x_2^2+(\e-it)^2)}}\right)^{\frac{1}{2}}\text{sgn}[-(x_1+x_2)],\nn\\
\sin\left(\frac{\theta_1-\theta_2}{2}\right)&=\left(\frac{1}{2}-\frac{x_1x_2-\e^2-t^2}{2\sqrt{(x_1^2+(\e+it)^2)}\sqrt{(x_2^2+(\e-it)^2)}}\right)^{\frac{1}{2}},\label{appA1,cos}
\]
where we define
\[
\text{sgn}[x]\equiv\label{sgnfunction}
\begin{cases}
1,&x>0,\\
-1,&x<0.
\end{cases}
\]
Substituting \eqref{appA1,cos} into \eqref{appA1,etabeforeac}, and let
\[
r_1=&\sqrt{x_1^2+(\e+it)^2},\quad r_2=\sqrt{x_2^2+(\e-it)^2},
\]
we find that the cross ratios can be expressed as
\begin{align}
\eta(x_1,x_2,t)=&\frac{(x_1+x_2+2t)+2\sqrt{(x_1+t)(x_2+t)+\epsilon^2+i\epsilon(x_1-x_2)}}{4\sqrt{(x_1+t)(x_2+t)+\epsilon^2+i\epsilon(x_1-x_2)}},\nonumber\\
\bar{\eta}(x_1,x_2,t)=&\frac{(x_1+x_2-2t)+2\sqrt{(x_1-t)(x_2-t)+\epsilon^2-i\epsilon(x_1-x_2)}}{4\sqrt{(x_1-t)(x_2-t)+\epsilon^2-i\epsilon(x_1-x_2)}}.\label{appA1,eta}
\end{align}
As $x_1=x_2=-l<0$, the above result coincides with that of \cite{Chen:2015usa}. According to \eqref{appA1,eta}, one can evaluate the early time limit and late time limits for cross ratios
\begin{align}
\lim_{t\to\infty}(\eta,\bar{\eta})\simeq&\bigg(1+\frac{(x_2-x_1+2i\epsilon)^2}{16t^2},-\frac{(x_2-x_1-2i\epsilon)^2}{16t^2}\bigg)\simeq(1,0),\label{app1:latetimeeta}\\
\lim_{t\to0}(\eta,\bar{\eta})\simeq&
\begin{cases}\Big(\frac{1}{2}+
\frac{x_1+x_2}{4\sqrt{x_1x_2}},\frac{1}{2}+
\frac{x_1+x_2}{4\sqrt{x_1x_2}}\Big),&x_1x_2>0,\\
\Big(\frac{1}{2}+\frac{x_1+x_2}{4\sqrt{x_1x_2}},\frac{1}{2}-\frac{x_1+x_2}{4\sqrt{x_1x_2}}\Big),&x_1>0>x_2,\\
\Big(\frac{1}{2}-\frac{x_1+x_2}{4\sqrt{x_1x_2}},\frac{1}{2}+\frac{x_1+x_2}{4\sqrt{x_1x_2}}\Big),&x_2>0>x_1.\label{early determin}
\end{cases}
\end{align}
\subsection{$A=[0,L]$}\label{appen-a2}
For the case of $A=[0, L]$, we can write the Euclidean cross ratios  in polar coordinates
\[
\eta=&\frac{1}{2}-\frac{(r_1r_4+r_2r_3)\cos\big(\frac{\theta_1-\theta_2-\theta_3+\theta_4}{2}\big)+i(r_1r_4-r_2r_3)\sin\big(\frac{\theta_1-\theta_2-\theta_3+\theta_4}{2}\big)}{4\sqrt{r_1}\sqrt{r_2}\sqrt{r_3}\sqrt{r_4}},\nn\\
\bar{\eta}=&\frac{1}{2}-\frac{(r_1r_4+r_2r_3)\cos\big(\frac{\theta_1-\theta_2-\theta_3+\theta_4}{2}\big)-i(r_1r_4-r_2r_3)\sin\big(\frac{\theta_1-\theta_2-\theta_3+\theta_4}{2}\big)}{4\sqrt{r_1}\sqrt{r_2}\sqrt{r_3}\sqrt{r_4}},\label{appA2,etabeforeac}
\]
where  $(r_1\cos\theta_1,r_{1}\sin\theta_1)=(x_1,-\t_{1})$, $(r_2\cos\theta_2,r_{2}\sin\theta_2)=(x_2,\t_{2})$, $(r_3\cos\theta_3,r_{3}\sin\theta_3)=(x_1-L,-\t_{1})$, $(r_4\cos\theta_4,r_{4}\sin\theta_4)=(x_2-L,\t_{2})$ $(0\leq\theta_{j}<2\pi,~j=1,2,3,4)$, and
\[
\cos(\theta_1-\theta_2)&=2\cos^2\left(\frac{\theta_1-\theta_2}{2}\right)-1=1-2\sin^2\left(\frac{\theta_1-\theta_2}{2}\right)=\frac{x_1x_2-\t_1\t_2}{\sqrt{(x_1^2+\t_1^2)}\sqrt{(x_2^2+\t_2^2)}},\nn\\
\cos(\theta_3-\theta_4)&=2\cos^2\left(\frac{\theta_3-\theta_4}{2}\right)-1=1-2\sin^2\left(\frac{\theta_3-\theta_4}{2}\right)=\frac{(x_1-L)(x_2-L)-\t_1\t_2}{\sqrt{(x_1-L)^2+\t_1^2}\sqrt{(x_2-L)^2+\t_2^2}}.
\]
Similar to the arguments in Appendix \ref{appen-a1} , we obtain the expressions of $\sin(\cos)\left(\frac{\theta_1-\theta_2}{2}\right)$ and $\sin(\cos)\left(\frac{\theta_3-\theta_4}{2}\right)$ after the analytic continuation as follows
\[
\cos\left(\frac{\theta_1-\theta_2}{2}\right)&=\left(\frac{1}{2}+\frac{x_1x_2-\e^2-t^2}{2\sqrt{(x_1^2+(\e+it)^2)}\sqrt{(x_2^2+(\e-it)^2)}}\right)^{\frac{1}{2}}\text{sgn}[-(x_1+x_2)],\nn\\
\sin\left(\frac{\theta_1-\theta_2}{2}\right)&=\left(\frac{1}{2}-\frac{x_1x_2-\e^2-t^2}{2\sqrt{(x_1^2+(\e+it)^2)}\sqrt{(x_2^2+(\e-it)^2)}}\right)^{\frac{1}{2}},\nn\\
\cos\left(\frac{\theta_3-\theta_4}{2}\right)&=\left(\frac{1}{2}+\frac{(x_1-L)(x_2-L)-\e^2-t^2}{2\sqrt{(x_1-L)^2+(\e+it)^2}\sqrt{(x_2-L)^2+(\e-it)^2}}\right)^{\frac{1}{2}}\text{sgn}[2L-(x_1+x_2)],\nn\\
\sin\left(\frac{\theta_3-\theta_4}{2}\right)&=\left(\frac{1}{2}-\frac{(x_1-L)(x_2-L)-\e^2-t^2}{2\sqrt{(x_1-L)^2+(\e+it)^2}\sqrt{(x_2-L)^2+(\e-it)^2}}\right)^{\frac{1}{2}},
\label{appA2,cos}
\]
where $\text{sgn}[x]$ is the sign function \eqref{sgnfunction} that we define.
Substituting \eqref{appA2,cos} into \eqref{appA2,etabeforeac}, and let
\begin{alignat}{2}
r_1=&\sqrt{x_1^2+(\e+it)^2},&\quad r_2=&\sqrt{x_2^2+(\e-it)^2},\nn\\
r_3=&\sqrt{(x_1-L)^2+(\e+it)^2},&\quad r_4=&\sqrt{(x_2-L)^2+(\e-it)^2},
\end{alignat}
we find that the cross ratios can be expressed as
\begin{align}
\eta(x_1,x_2,t)=&\frac{(x_1+x_2+2t)L\big[(L-x_1-t)(L-x_2-t)+\epsilon^2+i\epsilon(x_1-x_2)\Big]^{-\frac
{1}{2}}}{4\sqrt{(x_1+t)(x_2+t)+\epsilon^2+i\epsilon(x_1-x_2)}}\nn\\
&+\frac{1}{2}\left(1-\sqrt{\frac{(x_1+t)(x_2+t)+\epsilon^2+i\epsilon(x_1-x_2)}{(L-x_1-t)(L-x_2-t)+\epsilon^2+i\epsilon(x_1-x_2)}}\right),\nonumber\\
\bar{\eta}(x_1,x_2,t)=&\frac{(x_1+x_2-2t)L\big[(L-x_1+t)(L-x_2+t)+\epsilon^2-i\epsilon(x_1-x_2)\Big]^{-\frac
{1}{2}}}{4\sqrt{(x_1-t)(x_2-t)+\epsilon^2-i\epsilon(x_1-x_2)}}\nn\\
&+\frac{1}{2}\left(1-\sqrt{\frac{(x_1-t)(x_2-t)+\epsilon^2-i\epsilon(x_1-x_2)}{(L-x_1+t)(L-x_2+t)+\epsilon^2-i\epsilon(x_1-x_2)}}\right).\label{appA2,etabafterac}
\end{align}
As a self-consistent test, it can be found that \eqref{appA2,etabafterac} degenerates to \eqref{appA1,eta} as $L\to\infty$.
\section{Derivation of Eq.\eqref{resultsfinal}}\label{app2:derivation}
Let us first define a series of normalized excited states with the help of $\mathcal O_p$
\begin{align}
|\mathcal{O}_p(x)\rangle:=&\frac{1}{\sqrt{\langle\mathcal{O}_p^{\dagger}(x,\epsilon)\mathcal{O}_p(x,-\epsilon)\rangle}}\mathcal{O}_p(x,-\epsilon)|\Omega\rangle,\nn\\ \bigg(\langle\mathcal{O}_{p'}(x')|\mathcal{O}_{p}(x)\rangle=&\frac{\delta_{pp'}\cdot\langle\mathcal{O}_{p}^{\dagger}(x',\epsilon)\mathcal{O}_p(x,-\epsilon)\rangle}{\sqrt{\langle\mathcal{O}_p^{\dagger}(x,\epsilon)\mathcal{O}_p(x,-\epsilon)\rangle\langle\mathcal{O}_{p}^{\dagger}(x',\epsilon)\mathcal{O}_{p}(x',-\epsilon)\rangle}}\bigg).\label{Opstate}
\end{align}
$|\psi\rangle$ and $|\tilde\psi\rangle$ can be written as two superposition states of $|\mathcal{O}_p\rangle$
\[
|\psi\rangle=\sum_p\sqrt{\lambda_p}|\mathcal{O}_p(x)\rangle,\quad |\tilde\psi\rangle=\sum_p\sqrt{\tilde\lambda_p}|\mathcal{O}_p(\tilde x)\rangle,
\]
where
\begin{align}
\lambda_p=\frac{(C_p)^{2}{\langle\mathcal{O}_p^\dagger(x,\epsilon)\mathcal{O}_p(x,-\epsilon)\rangle}}{\sum_{p'}|C_{p'}|^2\langle\mathcal{O}_{p'}^\dagger(x,\epsilon)\mathcal{O}_{p'}(x,-\epsilon)\rangle},\quad\sum_p|\lambda_p|=1,\nonumber\\
\tilde\lambda_p=\frac{(\tilde C_p)^{2}{\langle\mathcal{O}_p^\dagger(\tilde x,\epsilon)\mathcal{O}_p(\tilde x,-\epsilon)\rangle}}{\sum_{p'}|\tilde C_{p'}|^2\langle\mathcal{O}_{p'}^\dagger(\tilde x,\epsilon)\mathcal{O}_{p'}(\tilde x,-\epsilon)\rangle},\quad\sum_p|\tilde\lambda_p|=1.\label{lambdap}
\end{align}
Generally speaking, $|\mathcal{O}_p\rangle$ is an entangled state living in the sub-Verma module $H_p\bigotimes H_{\bar{p}}$.  It can be written in the following form by Schmidt decomposition
\[
|\mathcal{O}_p(x)\rangle=\sum_ia^p_i(x)|p_i(x)\rangle\otimes|(\bar p_i(x)\rangle,
\]
where $\{|p_i(x)\rangle\}$ and $\{|\bar{p}_i(x)\rangle\}$ parameterized by $x$  are two orthonormal basises of $H_p$ and $H_{\bar{p}}$  repectively, and $a^p_i(x)$ are real coefficients. In these basisies the $n$th R\'enyi entropy of $|\mathcal{O}_p(x)\rangle$ reads
\[
S^{(n)}[\mathcal{O}_p(x)]=\frac{1}{1-n}\log\big\{\Tr_{(\oplus_p H_p)}\big[\left(\Tr_{(\oplus_p H_{\bar{p}})}|\mathcal{O}_p(x)\rangle\langle\mathcal{O}_p(x)|\right)^n\big]\big\}=\frac{1}{1-n}\log \sum_i\big(a^p_i(x)\big)^{2n}\label{compare1},
\]
meanwhile, the transition matrix becomes
\begin{align}
\mathcal{T}^{\psi|\tilde\psi}=\frac{1}{\sum_p\sqrt{\lambda_p}\sqrt{\tilde\lambda^*_p}\langle\mathcal{O}_p(\tilde x)|\mathcal{O}_p(x)\rangle}\sum_{p.p'}\sqrt{\lambda_p}\sqrt{\tilde\lambda^*_{p'}}\sum_{i,j}a^p_i(x)a^{p'}_j(\tilde x)|p_i(x)\rangle|\bar p_i(x)\rangle\langle p'_j(\tilde x)|\langle\bar p'_j(\tilde x)|.
\end{align}
The reduce transition matrix is obtained by tracing out the anti-holomorphic part,
\begin{align}
\mathcal{T}_H^{\psi|\tilde\psi}=&\Tr_{\oplus\bar pH_{\bar p}}\mathcal{T}^{\psi|\tilde\psi}\nonumber\\
=&\sum_{p}\sum_{i,j,k}\frac{\sqrt{\lambda_p}\sqrt{\tilde\lambda^*_{p}}a^p_i(x)a^{p}_j(\tilde x)\langle \bar p_j(\tilde x)|\bar p_{i}(x)\rangle\langle p_j(\tilde x)|p_k(x)\rangle}{\sum_{p''}\sqrt{\lambda_{p''}}\sqrt{\tilde\lambda^*_{p''}}\langle\mathcal{O}_{p''}(\tilde x)|\mathcal{O}_{p''}(x)\rangle}\cdot|p_i(x)\rangle\langle p_k(x)|\label{uneasy},
\end{align}
which in general is off-diagonal. We can compute the trace of $(\mathcal{T}_{H}^{\psi|\tilde\psi})^n$,
\begin{align}
&\Tr\left[(\mathcal{T}_{H}^{\psi|\tilde\psi})^n\right]\nn\\
=&\sum_{p}\frac{\Big(\sqrt{\lambda_p}\sqrt{\tilde\lambda^*_{p}}\Big)^n}{\Big(\sum_{p''}\sqrt{\lambda_{p''}}\sqrt{\tilde\lambda^*_{p''}}\langle\mathcal{O}_{p''}(\tilde x)|\mathcal{O}_{p''}(x)\rangle\Big)^n}\nonumber\\
&\times\sum_{\{i\},\{j\} }a^p_{i_1}(x)a^{p}_{j_1}(\tilde x)\langle \bar p_{j_1}(\tilde x)|\bar p_{i_1}(x)\rangle\langle p_{j_1}(\tilde x)|p_{i_2}(x)\rangle...a^p_{i_n}(x)a^{p}_{j_n}(\tilde x)\langle \bar p_{j_n}(\tilde x)|\bar p_{i_n}(x)\rangle\langle p_{j_n}(\tilde x)|p_{i_1}(x)\rangle\label{tobereduce}.
\end{align}
To further reduce $\eqref{tobereduce}$, let us turn to consider a more straightforward case that
\[
|\phi(t)\rangle=\text{e}^{-iHt}|\mathcal{O}_p(x)\rangle,\quad |\tilde\phi(t)\rangle=\text{e}^{-iHt}|\mathcal{O}_p(\tilde x)\rangle.
\]
According to the analysis in section \ref{sec4}, we know that the late time limit of $\Delta S^{(n)}\big[\mathcal{T}_A^{\phi|\tilde\phi}(t)\big]$ is equal to $\lim_{t\to\infty}\Delta S^{(n)}\big[\Tr_{A^c}|\phi(t)\rangle\langle\phi(t)|\big]$, and the latter we already know is equal to \eqref{compare1} \cite{Guo:2018lqq}. On the other hand, following the logic in \cite{Guo:2018lqq}, it's natural to expect that
\begin{align}
&\lim_{t\to\infty}\Delta S_A^{(n)}\big[\mathcal{T}^{\phi|\tilde\phi}(t)\big]\nn\\
=&\frac{1}{1-n}\log\Tr_{\oplus_pH_p}\Big[\big(\mathcal{T}_{H}^{\phi(0)|\tilde\phi(0)}\big)^n\Big]\nonumber\\
=&\frac{1}{1-n}\log\Big[\langle\mathcal{O}_p(\tilde x)|\mathcal{O}_p(x)\rangle^{-n}\nonumber\\
\times&{\sum_{\{i\},\{j\} }\alpha^p_{i_1}(x)\alpha^p_{j_1}(\tilde x)\langle \bar p_{j_1}(\tilde x)|\bar p_{i_1}(x)\rangle\langle p_{j_1}(\tilde x)|p_{i_2}(x)\rangle...\alpha^p_{i_n}(x)\alpha^p_{j_n}(\tilde x)\langle \bar p_{j_n}(\tilde x)|\bar p_{i_n}(x)\rangle\langle p_{j_n}(\tilde x)|p_{i_1}(x)\rangle}\Big]\label{compare2}
\end{align}
Comparing Eq.$\eqref{compare1}$ with Eq.$\eqref{compare2}$, we obtain the equality
\begin{align}
&{\sum_{\{i\},\{j\} }\alpha^p_{i_1}(x)\alpha^p_{j_1}(\tilde x)\langle \bar p_{j_1}(\tilde x)|\bar p_{i_1}(x)\rangle\langle p_{j_1}(\tilde x)|p_{i_2}(x)\rangle...\alpha^p_{i_n}(x)\alpha^p_{j_n}(\tilde x)\langle \bar p_{j_n}(\tilde x)|\bar p_{i_n}(x)\rangle\langle p_{j_n}(\tilde x)|p_{i_1}(x)\rangle}\nonumber\\
=&\langle\mathcal{O}_p(\tilde x)|\mathcal{O}_p(x) \rangle^n\sum_i\big(a^p_i(x)\big)^{2n}\label{equality}
\end{align}
Substituting Eq.$\eqref{equality}$ into Eq.$\eqref{tobereduce}$ and taking some algebra,  we finally obtain
\[
\frac{1}{1-n}\log\Tr\left[(\mathcal{T}_{H}^{\psi|\tilde\psi})^n\right]=\frac{1}{1-n}\log\left[\sum_p\left(\frac{C_p\tilde C_p^*\langle\mathcal{O}^\dagger_p(\tilde w,\bar {\tilde{w}})\mathcal{O}_p(w,\bar {{w}})\rangle}{\sum_{p'}C_{p'}\tilde C^*_{p'}\langle\mathcal{O}^\dagger_{p'}(\tilde w,\bar{\tilde{w}})\mathcal{O}_{p'}(w,\bar w)\rangle}\right)^n\text{e}^{(1-n)S^{(n)}[\mathcal{O}_p]}\right],\label{resultsfinalapp}
\]
which, in the light of the logic in \cite{Guo:2018lqq}, just corresponds to the late time limit of $n$th pseudo-R\'enyi entropy of $A$. Let $\{\tilde C_p\}=\{C_p\}=1$ and $\tilde x=x$, it can be readily found that $\eqref{resultsfinalapp}$ is reduced to the Eq.(2.26) in \cite{Guo:2018lqq}.
\bibliographystyle{JHEP}
\bibliography{bibforPseudoEntropy}{}
\end{document}